\documentclass[%
reprint,
preprintnumbers,
nofootinbib,
nobibnotes,
amsmath,amssymb,
aps,
prc,
]{revtex4-2}
\usepackage[pdftex]{graphicx}
\usepackage{dcolumn}
\usepackage{bm}
\usepackage[pdftex]{color}
\usepackage{CJK}
\usepackage[T1]{fontenc}
\usepackage{mathrsfs}
\def\ve#1{{\bm{#1}}}
\def\nuc#1#2#3{{}^{#2}_{#3}\mathrm{#1}}
\def\urm#1{\scriptstyle{\text{\textrm{\textmd{\textup{#1}}}}}}

\def\ca#1{{\mathcal{#1}}}

\let\temp\epsilon
\let\epsilon\varepsilon
\let\varepsilon\temp
\let\temp\relax
\let\temp\phi
\let\phi\varphi
\let\varphi\temp
\let\temp\relax
\DeclareMathOperator{\laplace}{\Delta}
\begin{document}
%
\begin{CJK*}{UTF8}{}
  \preprint{RIKEN-iTHEMS-Report-25}
  \title{Mirror-skin thickness: a possible observable sensitive to the charge symmetry breaking energy density functional}
  \author{Tomoya Naito (\CJKfamily{min}{内藤智也})}
  \affiliation{
    RIKEN Center for Interdisciplinary Theoretical and Mathematical Sciences (iTHEMS),
    Wako 351-0198, Japan}
  \affiliation{
    Department of Physics, Graduate School of Science, The University of Tokyo,
    Tokyo 113-0033, Japan}
  \author{Yuto Hijikata (\CJKfamily{min}{土方佑斗})}
  \affiliation{
    Department of Physics, Kyoto University,
    Kyoto 606-8502, Japan}
  \affiliation{
    RIKEN Nishina Center, Wako 351-0198, Japan}
  \author{Juzo Zenihiro (\CJKfamily{min}{銭廣十三})}
  \affiliation{
    Department of Physics, Kyoto University,
    Kyoto 606-8502, Japan}
  \affiliation{
    RIKEN Nishina Center, Wako 351-0198, Japan}
  \author{Gianluca Col\`{o}}
  \affiliation{
    Dipartimento di Fisica, Universit\`{a} degli Studi di Milano,
    Via Celoria 16, 20133 Milano, Italy}
  \affiliation{
    INFN, Sezione di Milano,
    Via Celoria 16, 20133 Milano, Italy}
  \author{Hiroyuki Sagawa (\CJKfamily{min}{佐川弘幸})}
  \affiliation{
    Center for Mathematics and Physics, University of Aizu,
    Aizu-Wakamatsu 965-8560, Japan}
  \affiliation{
    RIKEN Nishina Center, Wako 351-0198, Japan}
  \date{\today}
  \begin{abstract}
    We propose a new observable, named the mirror-skin thickness,
    in order to extract the strength of the charge symmetry breaking (CSB) term in an energy density functional (EDF).
    The mirror-skin thickness of $ N = 20 $ isotones and $ Z = 20 $ isotopes is studied by using Hartree-Fock-Bogoliubov (HFB) calculations with various  Skyrme EDFs and adding CSB and charge independence breaking (CIB) terms.
    It is shown that the mirror-skin thickness is sensitive only to the CSB EDF,
    but hardly depends on either the isospin symmetric part of the nuclear EDF or the CIB term.
    Therefore, this observable can be used to extract the magnitude of the CSB term in the EDF quantitatively,
    either from experimental data or \textit{ab initio} calculations.
    We have studied the accuracy in the mirror-skin thickness that is needed to extract sensible information.
    Our study may also help to understand the inconsistency between
    the strength of the phenomenological CSB and that extracted  from \textit{ab initio} calculations~[Naito \textit{et al.}~Nuovo.~Cim.~C~\textbf{47}, 52 (2024)].
    Among possible mirror pairs for experimental study, we propose the mirror-skin thickness between $ {}^{42} \mathrm{Ca} $ and $ {}^{42} \mathrm{Ti} $,
    which could be accessed in future experiments in RIBF and/or FRIB.
  \end{abstract}
  \maketitle
\end{CJK*}
%
%
\section{Introduction}
\par
The nuclear interaction has almost a complete isospin symmetry,
i.e., the proton-proton, neutron-neutron, and the isospin $ T = 1 $ channel of the proton-neutron interactions are almost the same~\cite{
  Heisenberg1932Z.Phys.77_1,
  Cassen1936Phys.Rev.50_846,
  Wigner1937Phys.Rev.51_106}.
Accordingly, it is known that properties of mirror nuclei are quite similar,
while the breaking of isospin symmetry has been paid attention to.
A famous example is the so-called Okamoto-Nolen-Schiffer anomaly~\cite{
  Okamoto1964Phys.Lett.11_150,
  Nolen1969Annu.Rev.Nucl.Sci.19_471}
of the mass difference of mirror nuclei;
the Coulomb interaction is not enough to describe the mass difference of mirror nuclei.
Many works including Refs.~\cite{
  Hatsuda1991Phys.Rev.Lett.66_2851,
  Auerbach1992Phys.Lett.B282_263,
  Suzuki1992Nucl.Phys.A536_141,
  Saito1994Phys.Lett.B335_17,
  Dong2018Phys.Rev.C97_021301,
  Sagawa2024Phys.Rev.C109_L011302}
have attempted to solve this problem.
\par
One of the possible solutions to solve the anomaly is
the inclusion of the isospin symmetry breaking (ISB) terms of the nuclear interaction.
They are a small part of the whole,
but their contribution to several nuclear properties has been discussed,
mainly the masses~\cite{
  Baczyk2016ActaPhys.Pol.BProc.Suppl.8_539,
  Baczyk2018Phys.Lett.B778_178,
  Baczyk2019J.Phys.G46_03LT01, 
  Dong2018Phys.Rev.C97_021301,
  Sagawa2024Phys.Rev.C109_L011302}
and
the isobaric analogue states~\cite{
  Roca-Maza2018Phys.Rev.Lett.120_202501}.
The energy difference in the level scheme of mirror nuclei~\cite{
  Hoff2020Nature580_52,
  algora2024isospinbreaking71kr71br}, the differences between the 
electromagnetic transition probabilities~\cite{  
  Wimmer2021Phys.Rev.Lett.126_072501},
as well as
the isovector density of $ N = Z $ nuclei~\cite{
Sagawa2022Phys.Lett.B829_137072}, 
have been proposed as signatures of isospin symmetry breaking.
Recently, we showed that the ISB terms of the nuclear interaction
also affect
the estimation of the slope parameter of the symmetry energy,
often called the $ L $ parameter~\cite{
  Naito2022Phys.Rev.C106_L061306,
  Naito2023Phys.Rev.C107_064302}.
Note that these effects depend on the strengths of the ISB terms of the nuclear interaction.
\par
The ISB terms can be divided into two classes:
the charge symmetry breaking (CSB) and the charge independence breaking (CIB) ones.
The CSB term corresponds to the difference between the proton-proton interaction and the neutron-neutron one,
while the CIB one corresponds to the difference between the like-particle interaction and the different-particle one,
i.e.,
\begin{subequations}
  \begin{align}
    V_{\urm{CSB}}
    & =
      V_{nn} - V_{pp}, \\
    V_{\urm{CIB}}
    & =
      \frac{V_{pp} + V_{nn}}{2}
      -
      V_{pn}^{T = 1} .
  \end{align}
\end{subequations}
The CSB interaction originates from the mass difference of protons and neutrons
and from the $ \rho $-$ \omega $ and $ \pi $-$ \eta $ mixings
in a meson-exchange representation of the nucleon-nucleon interaction~\cite{
  Henley1969IsospininNuclearPhysics_15,
  Coon1977Nucl.Phys.A287_381,
  Henley1979MesonsinNucleiVolumeI_405,
  Coon1982Phys.Rev.C26_562,
  Coon1987Phys.Rev.C36_2189,
  Miller1990Phys.Rep.194_1}.
In a quantum chromodynamics (QCD) picture, it originates from $ u $- and $ d $-quark mass difference and
a partial restoration of $ \overline{q} q $ condensation in the nuclear medium~\cite{
  Sagawa2024Phys.Rev.C109_L011302}.   
On the other hand, the CIB interaction mainly originates from the mass difference of charged pions and neutral ones~\cite{
  Henley1969IsospininNuclearPhysics_15,
  Henley1979MesonsinNucleiVolumeI_405,
  Miller1990Phys.Rep.194_1}.
There have been studies on ISB terms using the chiral effective field theory~\cite{
  Kolck1996Phys.Lett.B371_169,
  Friar1999Phys.Rev.C60_034006,
  Kolck2021Few-BodySyst.62_85}.
It has been found that the CSB terms give the dominant contribution to the isospin symmetry breaking of ground-state properties of atomic nuclei~\cite{
  Naito2023Phys.Rev.C107_064302};
thus, hereinafter, we focus on the CSB interaction only.
\par
The nuclear density functional theory (DFT)~\cite{
  Hohenberg1964Phys.Rev.136_B864,
  Kohn1965Phys.Rev.140_A1133,
  Vautherin1972Phys.Rev.C5_626}
is
a powerful tool to calculate nuclear properties of both ground and excited states systematically~\cite{
  Bogner2013Comput.Phys.Commun.184_2235,
  Colo2020Adv.Phys.X5_1740061}.
The starting point of the DFT calculation is an energy density functional (EDF).
To perform the DFT calculation with the ISB terms, 
it is indispensable to determine an ISB EDF.
\par
There are two ways to determine an EDF:
One is referring to \textit{ab initio} calculations and the other is referring to experimental data (phenomenological approach).
We have proposed a way to determine the CSB EDF using \textit{ab initio} calculations~\cite{
  Naito2022Phys.Rev.C105_L021304}
and using the QCD sum rule~\cite{
  Sagawa2024Phys.Rev.C109_L011302}.
However, the strengths obtained by these works are much smaller than the phenomenological ones.
This might be due to the fact that,
in the case of \textit{ab initio} calculations,
some nuclei display effects that cannot be accounted for so accurately like deformation or continuum effects.
It has to be stressed that ISB effects are smaller than other nuclear many-body correlations or,
in other words, they might be smaller than the theoretical accuracy of a number of methods that are available so far.
While this is not fully clarified and may deserve further study, at the same time,
a precise determination of ISB effects by using phenomenological approaches is still indispensable.
\par
References~\cite{
  Baczyk2018Phys.Lett.B778_178,
  Roca-Maza2018Phys.Rev.Lett.120_202501,
  Baczyk2019J.Phys.G46_03LT01}
proposed ISB EDFs
to reproduce various experimental data,
which give rather different strengths for the ISB terms starting either from \textit{ab initio} or QCD-based approaches.
The possible reason of such a deviation may originate from the fact that the ISB terms are quite small compared to the isospin symmetric terms,
and most observables also depend on both the isospin symmetric and ISB terms;
thus it is difficult to pin down quantitatively the ISB strength.
Therefore, it is important to find a physical observable sensitive only to the ISB terms.
The status of CSB EDFs is summarized in Ref.~\cite{
  Naito2024NuovoCim.C47_52}.
\par
In this paper, we propose a new physical observable called ``mirror-skin thickness'',
which is basically sensitive to the CSB term only.
We introduce the so-called $ C $-representation of the Skyrme-like CSB CSB interactions,
and their effects on the mirror-skin thickness is discussed term by term.
\par
This paper is organized as follows:
First, Sect.~\ref{sec:mirror_define} gives
the definition of the mirror-skin thickness
and its estimation using the liquid drop model.
Second, Sect.~\ref{sec:theoretical} gives the theoretical framework of the CSB interaction in nuclear DFT.
Then, Sect.~\ref{sec:calc} gives calculation results.
Finally, Sect.~\ref{sec:summary} is devoted to the summary.
%
%
\section{Mirror-skin thickness}
\label{sec:mirror_define}
\par
We propose a new observable called ``mirror-skin thickness'' and defined by 
\begin{equation}
  \label{eq:def_mirrorskin}
  \Delta R_{\urm{mirror}} \left( Z, N \right)
  \equiv
  R_p \left( Z, N \right)
  -
  R_n \left( N, Z \right),
\end{equation}
where $ R_p $ and $ R_n $ are the proton and neutron root-mean-square radii, respectively.
For an atomic nucleus with the proton and neutron numbers $ Z $ and $ N $,
the radius is denoted by $ R_i \left( Z, N \right) $
($ i = p $, $ n $),
and alternately $ R_i \left( N, Z \right) $ is for a nucleus with the proton and neutron numbers $ N $ and $ Z $.
This mirror-skin thickness is exactly zero if neither the Coulomb nor ISB terms of nuclear interaction is considered.
It should be noted that the mirror-skin thickness for the $ N = Z $ nuclei is nothing but the proton-skin thickness.
\par
Before we perform numerical calculations, 
we estimate the mirror-skin thickness using the liquid-drop model proposed by Myers and \'{S}wi\k{a}tecki~\cite{
  Myers1980Nucl.Phys.A336_267}.
The proton and neutron root-mean-square radii, respectively, read~\cite{
  Myers1980Nucl.Phys.A336_267}  
\begin{subequations}
  \begin{align}
    R_p
    & =
      \sqrt{\frac{3}{5}}
      \left[
      R
      -
      \frac{1}{2} t
      +
      \frac{5}{2} \frac{b^2}{R}
      +
      \frac{1}{35}
      \left( \frac{9}{2K_{\infty}} + \frac{1}{4J} \right)
      Z e^2
      \right],
      \label{eq:liquid_Rp} \\
    R_n
    & =
      \sqrt{\frac{3}{5}}
      \left[
      R
      +
      \frac{1}{2} t
      +
      \frac{5}{2} \frac{b^2}{R}
      +
      \frac{1}{35}
      \left( \frac{9}{2K_{\infty}} - \frac{1}{4J} \right)
      Z e^2    
      \right]
      \label{eq:liquid_Rn}
  \end{align}
\end{subequations}
with
\begin{subequations}
  \begin{align}
    R
    & =
      r_0 A^{1/3} \left( 1 + \overline{\epsilon} \right), \\ 
    t
    & =
      \frac{3}{2}
      r_0 
      \frac{JI - \frac{1}{12} c_1 Z A^{-1/3}}{Q \left( 1 + \frac{9}{4} \frac{J}{Q} A^{-1/3} \right)}, \\
    I
    & =
      \frac{N - Z}{A}, \\
    \overline{\epsilon}
    & =
      \frac{- 2 a_2 A^{-1/3} + L \overline{\delta}^2 + c_1 Z^2 A^{-4/3}}{K_{\infty}}, \\
    \overline{\delta}
    & =
      \frac{I + \frac{3}{16} \frac{c_1}{Q} Z A^{-2/3}}{1 + \frac{9}{4} \frac{J}{Q} A^{-1/3}}, \\
    c_1
    & =
      \frac{3e^2}{5 r_0}.
  \end{align}
\end{subequations}
Here,
$ J \simeq 30 \, \mathrm{MeV} $~\cite{
  Roca-Maza2018Prog.Part.Nucl.Phys.101_96}
is the symmetry energy at the saturation density,
$ L $ is the slope of the symmetry energy,
$ K_{\infty} \simeq 226 \, \mathrm{MeV} $~\cite{
  Li2023Phys.Rev.Lett.131_082501}
is the incompressibility,
$ Q \simeq 17 \, \mathrm{MeV} $~\cite{
  Myers1980Nucl.Phys.A336_267}
is the effective surface stiffness coefficient,
$ a_2 \simeq 20 \, \mathrm{MeV} $~\cite{
  Myers1980Nucl.Phys.A336_267}
is the surface energy coefficient, 
$ b \simeq 1 \, \mathrm{fm} $ is the surface width~\cite{
  Warda2009Phys.Rev.C80_024316,
  Centelles2010Phys.Rev.C82_054314},
and $ r_0 \simeq 1.18 \, \mathrm{fm} $~\cite{  
  Myers1980Nucl.Phys.A336_267}.
Note that $ b $ can be different for protons and neutrons,
but for simplicity, we approximate that $ b $ for protons is identical to that for neutrons.
In Ref.~\cite{
  Warda2009Phys.Rev.C80_024316},
it was confirmed that
$ \sqrt{\frac{3}{5}} \frac{5}{2R} \left( b_n^2 - b_p^2 \right) $
is less than $ 0.1 \, \mathrm{fm} $ even in the case of large proton-neutron asymmetry, 
and the value is model independent.
Hence, this approximation is good enough in the current discussion. 
It should also be noted that this representation does not include the effect of isospin symmetry breaking other than the Coulomb interaction.
\par
We take the approximation
\begin{align}
  \frac{b^2}{R}
  & =
    \frac{b^2}{r_0 A^{1/3} \left( 1 + \overline{\epsilon} \right)}
    \notag \\
  & \simeq 
    \frac{b^2}{r_0 A^{1/3}}
    \left( 1 - \overline{\epsilon} \right)
\end{align}
since $ \overline{\delta} \simeq -0.07 $ and $ \overline{\epsilon} \simeq -0.05 $ (for $ \nuc{Ni}{48}{} $).
Then, we obtain
\begin{widetext}
  \begin{subequations}
    \begin{align}
      R_p \left( Z, N \right)
      & \simeq
        \sqrt{\frac{3}{5}}
        \left\{
        R_0 \left( Z, N \right)
        +
        R_1 \left( Z, N \right)
        +
        \left[
        \frac{9}{70 K_{\infty}}
        +
        R_2 \left( Z, N \right)
        +
        R_3 \left( Z, N \right)
        \right]
        Z e^2
        +
        R_4 \left( Z, N \right)
        Z^2 e^4
        \right\},
        \label{eq_LDM_Rp} \\
      R_n \left( Z, N \right)
      & \simeq
        \sqrt{\frac{3}{5}}
        \left\{
        R_0 \left( Z, N \right)
        -
        R_1 \left( Z, N \right)
        +
        \left[
        \frac{9}{70 K_{\infty}}
        +
        R_2 \left( Z, N \right)
        -
        R_3 \left( Z, N \right)
        \right]
        Z e^2
        +
        R_4 \left( Z, N \right)
        Z^2 e^4
        \right\}
        \label{eq_LDM_Rn} 
    \end{align}
  \end{subequations}
  with
  \begin{subequations}
    \begin{align}
      R_0 \left( Z, N \right)
      & =
        r_0 A^{1/3}
        +
        \frac{5}{2}
        \frac{b^2}{r_0 A^{1/3}}
        -
        \left(
        r_0 A^{1/3}
        -
        \frac{5}{2}
        \frac{b^2}{r_0 A^{1/3}}
        \right)
        \frac{1}{K_{\infty}}
        \left[
        2 a_2 A^{-1/3}
        -
        \frac{I^2 L}{\left( 1 + \frac{9}{4} \frac{J}{Q} A^{-1/3} \right)^2}
        \right], \\
      R_1 \left( Z, N \right)
      & = 
        -
        \frac{3}{4}
        r_0
        \frac{JI}{Q \left( 1 + \frac{9}{4} \frac{J}{Q} A^{-1/3} \right)}, \\
      R_2 \left( Z, N \right)
      & =
        \frac{1}{K_{\infty}}
        \left(
        r_0 A^{1/3}
        -
        \frac{5}{2}
        \frac{b^2}{r_0 A^{1/3}}
        \right)
        \left[
        Z A^{-4/3}
        +
        \frac{3}{8}
        \frac{I A^{-2/3}}{\left( 1 + \frac{9}{4} \frac{J}{Q} A^{-1/3} \right)^2}
        \frac{L}{Q}
        \right]
        \frac{3}{5 r_0}, \\
      R_3 \left( Z, N \right)
      & =
        \frac{3}{80}
        \frac{A^{-1/3}}{Q}
        +
        \frac{1}{140 J}
        -
        \frac{27}{320}
        \frac{J A^{-2/3}}{\left( 1 + \frac{9}{4} \frac{J}{Q} A^{-1/3} \right)}
        \frac{1}{Q^2}, \\
      R_4 \left( Z, N \right)
      & = 
        \frac{81}{800}
        \frac{1}{K_{\infty}}
        \frac{L \left( 2 r_0^2 A^{2/3} - 5 b^2 \right)}{16 r_0^3 A^{5/3}}
        \frac{1}{\left( 1 + \frac{9}{4} \frac{J}{Q} A^{-1/3} \right)^2}
        \frac{1}{Q^2},
    \end{align}
  \end{subequations}
  where
  $ R_0 \left( Z, N \right) = R_0 \left( N, Z \right) $,
  $ R_1 \left( Z, N \right) = - R_1 \left( N, Z \right) $,
  $ R_3 \left( Z, N \right) = R_3 \left( N, Z \right) $,
  and
  $ R_4 \left( Z, N \right) = R_4 \left( N, Z \right) $
  hold.
  Therefore, the mirror-skin thickness in the liquid-drop model reads
  \begin{align}
    & \Delta R_{\urm{mirror}} \left( Z, N \right)
      \notag \\
    & =
      R_p \left( Z, N \right)
      -
      R_n \left( Z, N \right)
      \notag \\
    & \simeq
      \sqrt{\frac{3}{5}}
      \left\{
      R_0 \left( Z, N \right)
      +
      R_1 \left( Z, N \right)
      +
      \left[
      \frac{9}{70 K_{\infty}}
      +
      R_2 \left( Z, N \right)
      +
      R_3 \left( Z, N \right)
      \right]
      Z e^2
      +
      R_4 \left( Z, N \right)
      Z^2 e^4
      \right\}
      \notag \\
    & \quad
      -
      \sqrt{\frac{3}{5}}
      \left\{
      R_0 \left( N, Z \right)
      -
      R_1 \left( N, Z \right)
      +
      \left[
      \frac{9}{70 K_{\infty}}
      +
      R_2 \left( N, Z \right)
      -
      R_3 \left( N, Z \right)
      \right]
      N e^2
      +
      R_4 \left( N, Z \right)
      N^2 e^4
      \right\}
      \notag \\
    & =
      \sqrt{\frac{3}{5}}
      \left\{
      \left( Z - N \right) 
      \frac{9}{70 K_{\infty}}
      +
      \left[
      Z R_2 \left( Z, N \right)
      -
      N R_2 \left( N, Z \right)
      \right]
      +
      A R_3
      +
      R_4 \left( Z, N \right)
      \left( Z^2 - N^2 \right) e^2
      \right\}
      e^2
      \notag \\
    & =
      \sqrt{\frac{3}{5}}
      \left(
      \left( Z - N \right)
      \left\{
      \frac{9}{70 K_{\infty}}
      +
      \frac{1}{K_{\infty}}
      \left(
      r_0 A^{1/3}
      -
      \frac{5}{2}
      \frac{b^2}{r_0 A^{1/3}}
      \right)
      \left[
      A^{-1/3}
      -
      \frac{3}{8}
      \frac{A^{-2/3}}{\left( 1 + \frac{9}{4} \frac{J}{Q} A^{-1/3} \right)^2}
      \frac{L}{Q}
      \right]
      \frac{3}{5 r_0}
      \right.
      \right.
      \notag \\
    & \qquad
      \left.
      \left.
      \vphantom{
      \frac{9}{70 K_{\infty}}
      +
      \frac{1}{K_{\infty}}
      \left(
      r_0 A^{1/3}
      -
      \frac{5}{2}
      \frac{b^2}{r_0 A^{1/3}}
      \right)
      \left[
      A^{-1/3}
      -
      \frac{3}{8}
      \frac{A^{-2/3}}{\left( 1 + \frac{9}{4} \frac{J}{Q} A^{-1/3} \right)^2}
      \frac{L}{Q}
      \right]
      \frac{3}{5 r_0}}
      +
      A
      R_4 \left( Z, N \right)
      e^2
      \right\}
      +
      A R_3 \left( Z, N \right)
      \right)
      e^2.
      \label{eq:LDM_Rmirror}
  \end{align}
\end{widetext}
The mirror-skin thickness is zero 
if we neglect the Coulomb interaction, i.e., $ e^2 = 0 $. 
\par
For $ \nuc{Ni}{48}{} $, the mirror-skin thickness in the liquid-drop model [Eq.~\eqref{eq:LDM_Rmirror}] reads 
\begin{align}
  \Delta R_{\urm{mirror}} \left( {Z = 28}, {N = 20} \right)
  & =
    0.0028 \left( Z - N \right)
    +
    0.00059 A
    \, \mathrm{fm}
    \notag \\
  & =
    0.051 \, \mathrm{fm},
\end{align}
We can easily find that 
even in $ N = Z $ nuclei, the mirror-skin thickness, i.e., the proton-skin thickness, has a finite value.
The expression for $ \Delta R_{\urm{mirror}} $ in the liquid-drop model in Eq.~\eqref{eq:LDM_Rmirror} shows that
a proton-rich nucleus ($ Z > N $) has a larger $ \Delta R_{\urm{mirror}} $ than the neutron-rich one.
We should also notice that $ \Delta R_{\urm{mirror}} $ could be negative for $ N \gg Z $ nuclei,
while it will be shown that all the DFT calculation always gives positive $ \Delta R_{\urm{mirror}} $.
%
%
\section{Theoretical framework}
\label{sec:theoretical}
\subsection{Calculation setup}
\par
We perform Skyrme Hartree-Fock-Bogoliubov (HFB) calculations~\cite{
  Vautherin1972Phys.Rev.C5_626,
  Dobaczewski1984Nucl.Phys.A422_103}
by assuming the spherical symmetry.
The radial wave function 
is obtained from the HFB solution in coordinate space, with a spatial mesh
of $ 0.1 \, \mathrm{fm} $ and a box size of $ 16 \, \mathrm{fm} $.
\par
The SLy4 EDF~\cite{
  Chabanat1998Nucl.Phys.A635_231}
and 
the volume-type pairing interaction~\cite{
  Dobaczewski1984Nucl.Phys.A422_103}
are used for the isospin symmetric particle-hole and particle-particle channel, respectively.
The Hartree-Fock-Slater approximation, i.e., the local density approximation,
is used for the Coulomb interaction~\cite{
  Slater1951Phys.Rev.81_385}~\footnote{
  We have confirmed that the way of modelling of the Coulomb interaction does not affect proton or neutron radii.
  The detail is shown in Appendix~\ref{sec:Coul}.}.
The single-particle orbitals up to $ 60 \, \mathrm{MeV} $ in the Hartree-Fock equivalent energy~\cite{
  Stoitsov2013Comput.Phys.Commun.184_1592}
is considered for the HFB calculation
and 
the strength of the pairing interaction
($ -194.2 \, \mathrm{MeV} \, \mathrm{fm}^3 $)
is determined to reproduce the neutron pairing gap of $ \nuc{Sn}{120}{} $
as $ 1.4 \, \mathrm{MeV} $~\cite{
  Naito2023Phys.Rev.C107_054307}.
The proton-proton pairing strength is set to the same as the neutron-neutron one~\footnote{
  It has been discussed that the Coulomb repulsion between protons weakens the proton-proton pairing strength~\cite{
    Anguiano2001Nucl.Phys.A683_227,
    Hilaire2002Phys.Lett.B531_61,
    Nakada2011Phys.Rev.C83_031302},
  while we have confirmed that its effect is the order of $ 0.001 \, \mathrm{fm} $.}.
On top of it, the CSB EDF is considered, which will be explained below.
Note that the pairing  strength is kept even if the ISB interaction is considered.
\par
We focus on
$ Z = 20 $ isotopes and $ N = 20 $ isotones 
with the mass number $ A = 38 $, $ 40 $, $ 42 $, $ 44 $, $ 46 $, and $ 48 $
in our numerical study of the mirror-skin thickness
since they are proton- or neutron-magic nuclei and thus expected to be spherical.
\subsection{Skyrme-type CSB interaction}
\par
The Skyrme-type CSB interaction is defined by~\cite{
  Roca-Maza2018Phys.Rev.Lett.120_202501,
  Baczyk2018Phys.Lett.B778_178,
  Baczyk2019J.Phys.G46_03LT01,
  Sagawa2019Eur.Phys.J.A55_227,
  Naito2023Phys.Rev.C107_064302}
\begin{widetext}
  \begin{equation}
    \label{eq:Skyrme_CSB} 
    v_{\urm{Sky}}^{\urm{CSB}} \left( \ve{r} \right)
    =
    \left\{
      s_0
      \left( 1 + y_0 P_{\sigma} \right)
      \delta \left( \ve{r} \right)
      +
      \frac{s_1}{2}
      \left( 1 + y_1 P_{\sigma} \right)
      \left[
        \ve{k}^{\dagger 2} \delta \left( \ve{r} \right)
        +
        \delta \left( \ve{r} \right) \ve{k}^2
      \right]
      +
      s_2
      \left( 1 + y_2 P_{\sigma} \right)
      \ve{k}^{\dagger}
      \cdot
      \delta \left( \ve{r} \right)
      \ve{k}
    \right\}
    \frac{\tau_{z1} + \tau_{z2}}{4},
  \end{equation}
  where
  $ \ve{r} = \ve{r}_1 - \ve{r}_2 $,
  $ \ve{R} = \left( \ve{r}_1 + \ve{r}_2 \right) / 2 $,
  and
  $ P_{\sigma} = \left( 1 + \ve{\sigma}_1 \cdot \ve{\sigma}_2 \right) / 2 $.
  Here, $ \tau_{zj} $ is the $ z $-projection of the isospin operator for the nucleon $ j $
  ($ \tau_{zj} = +1 $ for neutrons and $ \tau_{zj} = -1 $ for protons)
  and $ \ve{k} $ is the operator of the relative momentum.
  Accordingly the CSB energy density reads~\cite{
    Sagawa2019Eur.Phys.J.A55_227,
    Naito2023Phys.Rev.C107_064302}
  \begin{equation}
    \label{eq:CSB_EDF-tx} 
    \ca{E}_{\urm{CSB}}
    =
    \frac{\tilde{s}_0}{8}
    \left( \rho_n^2 - \rho_p^2 \right)
    +
    \frac{1}{16}
    \left(
      \tilde{s}_1 
      +
      3 \tilde{s}_2 
    \right)
    \left( \rho_n \tau_n - \rho_p \tau_p \right)
    -
    \frac{3}{64}
    \left(
      \tilde{s}_1 
      -
      \tilde{s}_2 
    \right)
    \left( \rho_n \laplace \rho_n - \rho_p \laplace \rho_p \right)
    -
    \frac{1}{32}
    \left(
      \tilde{s}_1 
      -
      \tilde{s}_2 
    \right)
    \left( \ve{J}_n^2 - \ve{J}_p^2 \right),
  \end{equation}
\end{widetext}
where $ \rho $, $ \tau $, and $ \ve{J} $ are the particle, kinetic, and spin-orbit densities, respectively.
Since the parameters $ s_j $ and $ y_j $ always appear together as 
$ s_0 \left( 1 - y_0 \right) $,
$ s_1 \left( 1 - y_1 \right) $,
and
$ s_2 \left( 1 + y_2 \right) $,
we define
\begin{subequations}
  \begin{align}
    \tilde{s}_0
    & =
      s_0 \left( 1 - y_0 \right), \\
    \tilde{s}_1
    & = 
      s_1 \left( 1 - y_1 \right), \\
    \tilde{s}_2
    & = 
      s_2 \left( 1 + y_2 \right).
  \end{align}
\end{subequations}
Hereinafter, the $ \ve{J}^2 $ term is neglected for simplicity, as did in the SLy4 EDF~\cite{
  Chabanat1998Nucl.Phys.A635_231}.
\par
The so-called $ C $-representation of a Skyrme EDF~\cite{
  Dobaczewski1995Phys.Rev.C52_1827}
is also widely used.
This $ C $-representation adopts the coupling constants associated with the different types of densities, generalized densities or its gradients,
which is convenient in the present context.
We define the CSB EDF in the $ C $-representation as
\begin{widetext}
  \begin{equation}
    \label{eq:CSB_EDF-C} 
    \ca{E}_{\urm{CSB}}
    =
    C_{\urm{CSB}}^{\rho}
    \left( \rho_n^2 - \rho_p^2 \right)
    +
    C_{\urm{CSB}}^{\tau}
    \left( \rho_n \tau_n - \rho_p \tau_p \right)
    + 
    C_{\urm{CSB}}^{\laplace \rho}
    \left( \rho_n \laplace \rho_n - \rho_p \laplace \rho_p \right),
  \end{equation}
\end{widetext}
where these $ C $-parameters are given by
\begin{subequations}
  \begin{align}
    C_{\urm{CSB}}^{\rho}
    & = 
      \frac{\tilde{s}_0}{8}, \\
    C_{\urm{CSB}}^{\tau}
    & = 
      \frac{1}{16}
      \left(
      \tilde{s}_1 
      +
      3 \tilde{s}_2 
      \right), \\
    C_{\urm{CSB}}^{\laplace \rho}
    & = 
      -
      \frac{3}{64}
      \left(
      \tilde{s}_1 
      -
      \tilde{s}_2 
      \right).
  \end{align}
\end{subequations}
Inversely, $ \tilde{s}_0 $, $ \tilde{s}_1 $, and $ \tilde{s}_2 $ are written as
\begin{subequations}
  \begin{align}
    \tilde{s}_0
    & =
      8
      C_{\urm{CSB}}^{\rho}, \\
    \tilde{s}_1
    & =
      4
      C_{\urm{CSB}}^{\tau}
      -
      16
      C_{\urm{CSB}}^{\laplace \rho}, \\
    \tilde{s}_2
    & = 
      4
      C_{\urm{CSB}}^{\tau}
      +
      \frac{16}{3}
      C_{\urm{CSB}}^{\laplace \rho},
  \end{align}
\end{subequations}
respectively.
%
%
\section{Calculation results}
\label{sec:calc}
\par
\subsection{EDF dependence of mirror-skin thickness}
\label{sec:calc_normal}
\par
\begin{figure}[tb]
  \centering
  \includegraphics[width=1.0\linewidth]{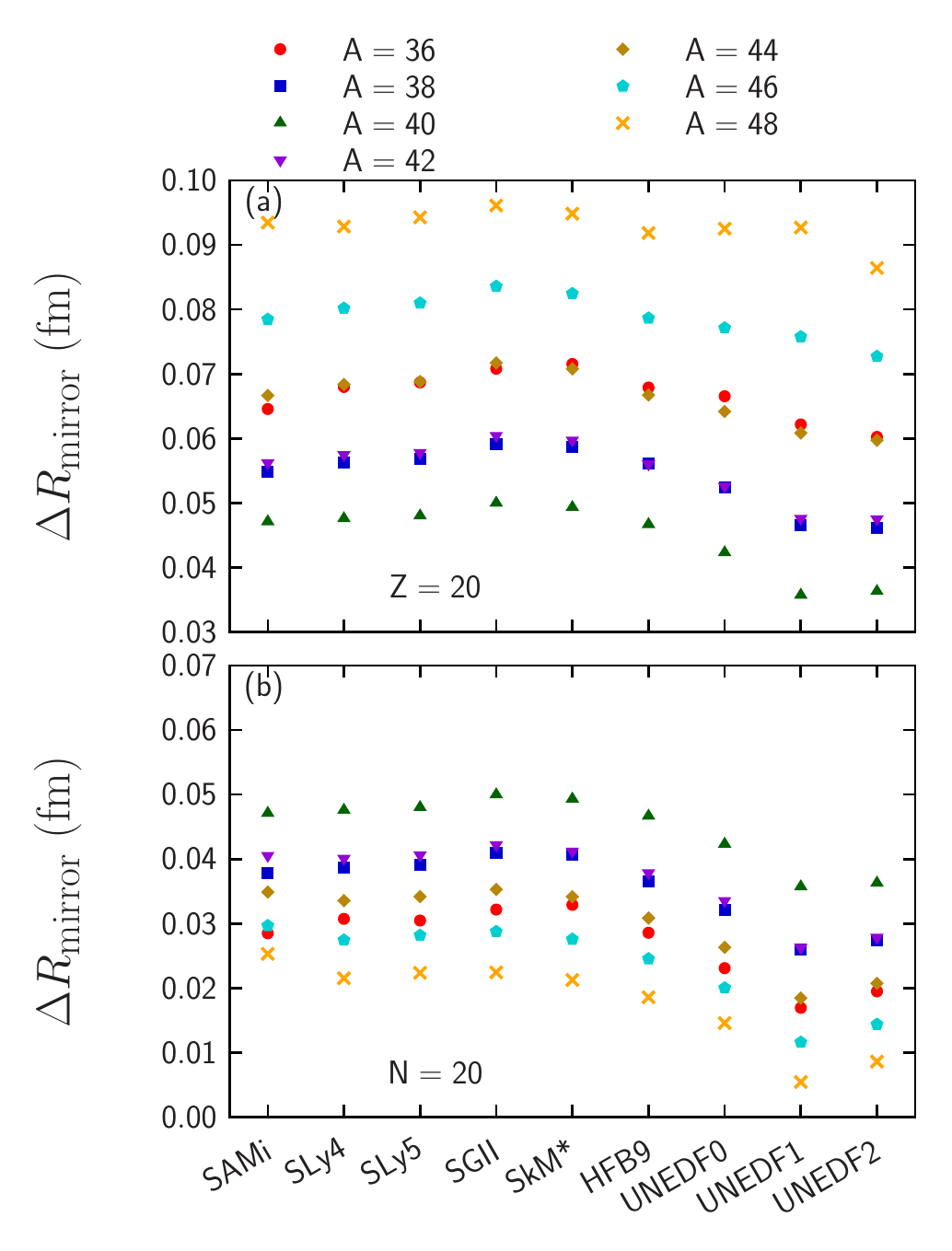}
  \caption{EDF dependence of the mirror-skin thickness $ \Delta R_{\urm{mirror}} $.
    The 
    SLy4, SLy5~\cite{
      Chabanat1998Nucl.Phys.A635_231},
    SAMi~\cite{
      Roca-Maza2012Phys.Rev.C86_031306},
    SGII~\cite{
      VanGiai1981Phys.Lett.B106_379},
    SkM*~\cite{
      Bartel1982Nucl.Phys.A386_79},
    HFB9~\cite{
      Goriely2005Nucl.Phys.A750_425},
    UNEDF0~\cite{
      Kortelainen2010Phys.Rev.C82_024313},
    UNEDF1~\cite{
      Kortelainen2012Phys.Rev.C85_024304},
    and UNEDF2~\cite{
      Kortelainen2014Phys.Rev.C89_054314}
    EDFs
    and 
    the volume-type pairing interaction~\cite{
      Dobaczewski1984Nucl.Phys.A422_103}
    are used for the isospin symmetric particle-hole and particle-particle channel, respectively.
    The Hartree-Fock-Slater approximation, i.e., the local density approximation,
    is used for the Coulomb interaction~\cite{
      Slater1951Phys.Rev.81_385}.}
  \label{fig:normal_Rmirror}
\end{figure}

Figure~\ref{fig:normal_Rmirror} shows the mirror-skin thickness calculated by different Skyrme EDFs:
SLy4, SLy5~\cite{
  Chabanat1998Nucl.Phys.A635_231},
SAMi~\cite{
  Roca-Maza2012Phys.Rev.C86_031306},
SGII~\cite{
  VanGiai1981Phys.Lett.B106_379},
SkM*~\cite{
  Bartel1982Nucl.Phys.A386_79},
HFB9~\cite{
  Goriely2005Nucl.Phys.A750_425},
UNEDF0~\cite{
  Kortelainen2010Phys.Rev.C82_024313},
UNEDF1~\cite{
  Kortelainen2012Phys.Rev.C85_024304},
and UNEDF2~\cite{
  Kortelainen2014Phys.Rev.C89_054314}.
These calculations using a variety of EDFs allow estimating the model dependence of the mirror-skin thickness. 
This is rather small, as shown in Fig.~\ref{fig:normal_Rmirror};
in fact, the variation among different EDFs is about $ 0.005 \, \mathrm{fm} $,
except when including the three variants of UNEDF: then, it can arrive at about $ 0.01 \, \mathrm{fm} $.
\par
\begin{figure}[tb]
  \centering
  \includegraphics[width=1.0\linewidth]{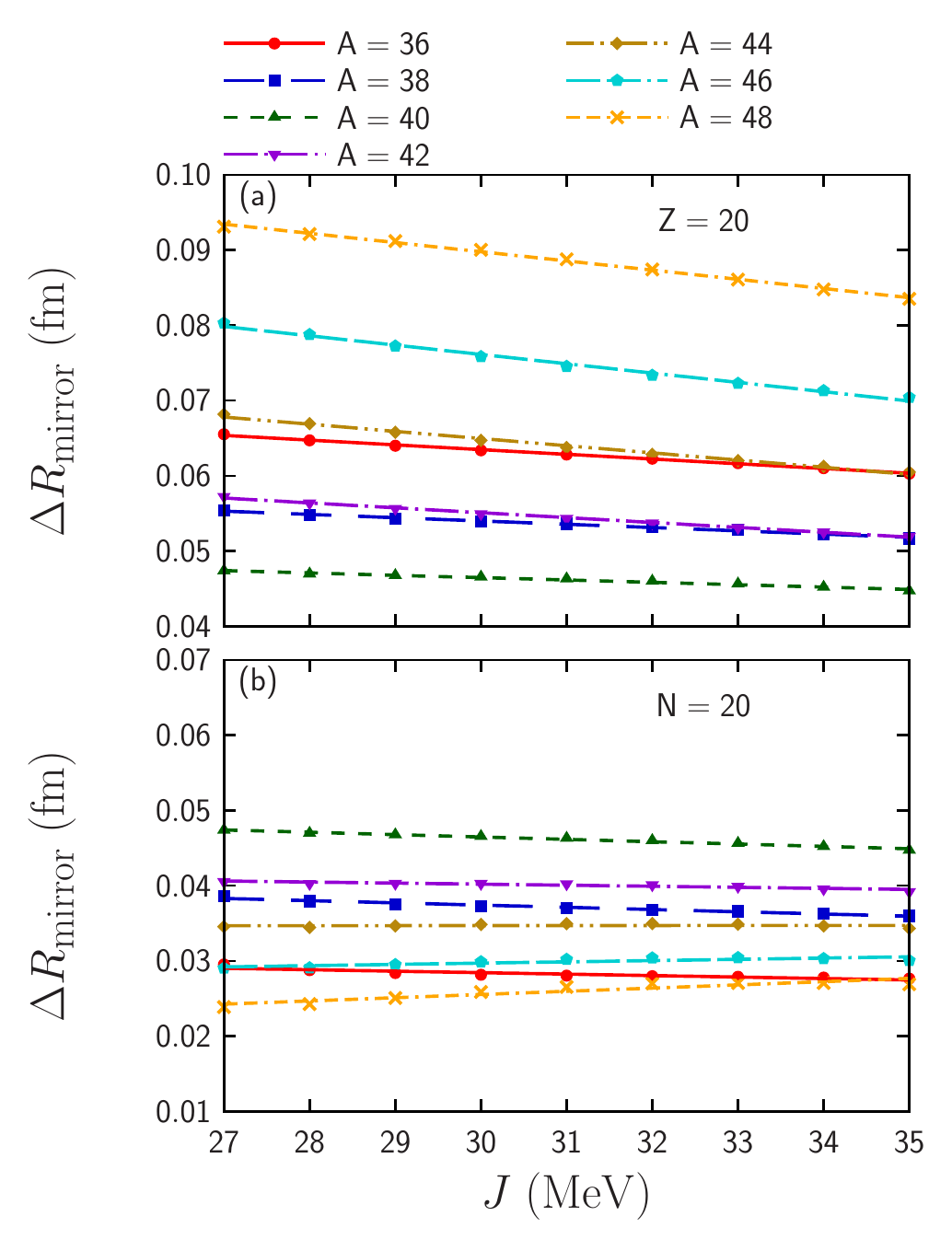}
  \caption{Mirror-skin thickness $ \Delta R_{\urm{mirror}} $ as functions of the symmetry energy at the saturation density $ J $.
    The SAMi-J family~\cite{
      Roca-Maza2013Phys.Rev.C87_034301}
    and 
    the volume-type pairing interaction~\cite{
      Dobaczewski1984Nucl.Phys.A422_103}
    are used for the isospin symmetric particle-hole and particle-particle channel, respectively.
    The Hartree-Fock-Slater approximation, i.e., the local density approximation,
    is used for the Coulomb interaction~\cite{
      Slater1951Phys.Rev.81_385}.}
  \label{fig:SAMiJ_Rmirror}
\end{figure}

Moreover, the mirror-skin thickness hardly depends on the symmetry energy.
Figure~\ref{fig:SAMiJ_Rmirror} shows the dependence of $ \Delta R_{\urm{mirror}} $ on the symmetry energy at saturation $ J $.
All these calculations are performed using the SAMi-J family~\cite{
  Roca-Maza2013Phys.Rev.C87_034301}.
The $ J $-dependence is much weaker than that on the CSB strength, as will be seen in Sect.~\ref{sec:calc_CSB}.
In particular, the symmetry energy dependence of the mirror-skin thickness in the $ N = 20 $ isotones is negligibly small.
\subsection{CIB dependence of mirror-skin thickness}
\label{sec:calc_CIB}
\par
We also check that the mirror-skin thickness does not depend on the CIB strength. 
The sensitivity with respect to the leading-order CIB strength
($ u_0 $ in the next formula, analogous to the $ t_0 $ in the standard Skyrme interaction)
is two orders of magnitude smaller than that to the CSB strength, which will be shown later.
Therefore, the mirror-skin thickness could be a useful quantity for pinning down the strength of the CSB interaction.
\par
\begin{figure}[tb]
  \centering
  \includegraphics[width=1.0\linewidth]{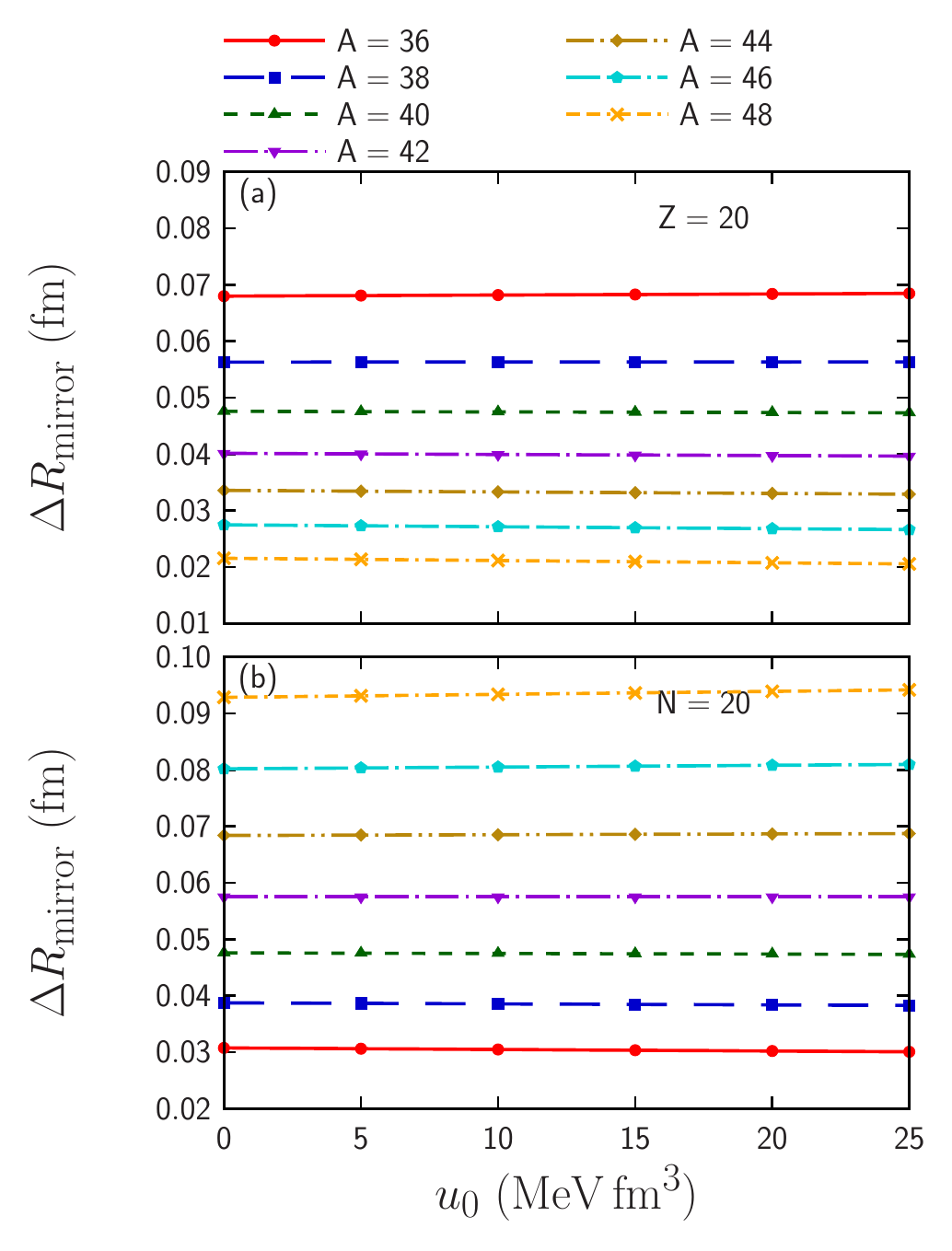}
  \caption{Mirror-skin thickness $ \Delta R_{\urm{mirror}} $ as functions of the CIB strength $ u_0 $.
    The SLy4 EDF~\cite{
      Chabanat1998Nucl.Phys.A635_231}
    and 
    the volume-type pairing interaction~\cite{
      Dobaczewski1984Nucl.Phys.A422_103}
    are used for the isospin symmetric particle-hole and particle-particle channel, respectively.
    The Hartree-Fock-Slater approximation, i.e., the local density approximation,
    is used for the Coulomb interaction~\cite{
      Slater1951Phys.Rev.81_385}.}
  \label{fig:u0_Rmirror}
\end{figure}

The leading-order Skyrme-like CIB interaction
\begin{equation}
  v_{\urm{Sky}}^{\urm{CIB}} \left( \ve{r} \right)
  =
  \frac{u_0}{2}
  \left( 1 + z_0 P_{\sigma} \right)
  \delta \left( \ve{r} \right)
  \tau_{z1} \tau_{z2}
\end{equation}
is considered on top of the SLy4 interaction and the volume-type pairing interaction.
The CIB strength is varied from $ u_0 = 0 $ to $ 25 \, \mathrm{MeV} \, \mathrm{fm}^3 $, taking $ z_0 = -1 $.
The mirror-skin thickness with the different CIB strength is shown in Fig.~\ref{fig:u0_Rmirror}.
It is seen that the mirror-skin thickness does not depend on the CIB interaction.
\subsection{CSB dependence of mirror-skin thickness}
\label{sec:calc_CSB}
\subsubsection{$ C_{\urm{CSB}}^{\rho} $-term dependence}
\par
First, we focus on the $ C_{\urm{CSB}}^{\rho} $ term in Eq.~\eqref{eq:CSB_EDF-C},
which corresponds to the $ s_0 $ term in Eq.~\eqref{eq:CSB_EDF-tx}.
Here, $ - C_{\urm{CSB}}^{\rho} $ is varied from $ 0 $ to $ 10 \, \mathrm{MeV} \, \mathrm{fm}^3 $,
which corresponds to $ 0 $ to $ 80 \, \mathrm{MeV} \, \mathrm{fm}^3 $ in $ - \tilde{s}_0 $,
considering neither the $ C_{\urm{CSB}}^{\tau} $ nor $ C_{\urm{CSB}}^{\laplace \rho} $ term.
Note that the strongest phenomenological CSB EDF has $ \tilde{s}_0 \simeq -50 \, \mathrm{MeV} \, \mathrm{fm}^3 $~\cite{
  Naito2024NuovoCim.C47_52}.
\par
\begin{figure}[tb]
  \centering
  \includegraphics[width=1.0\linewidth]{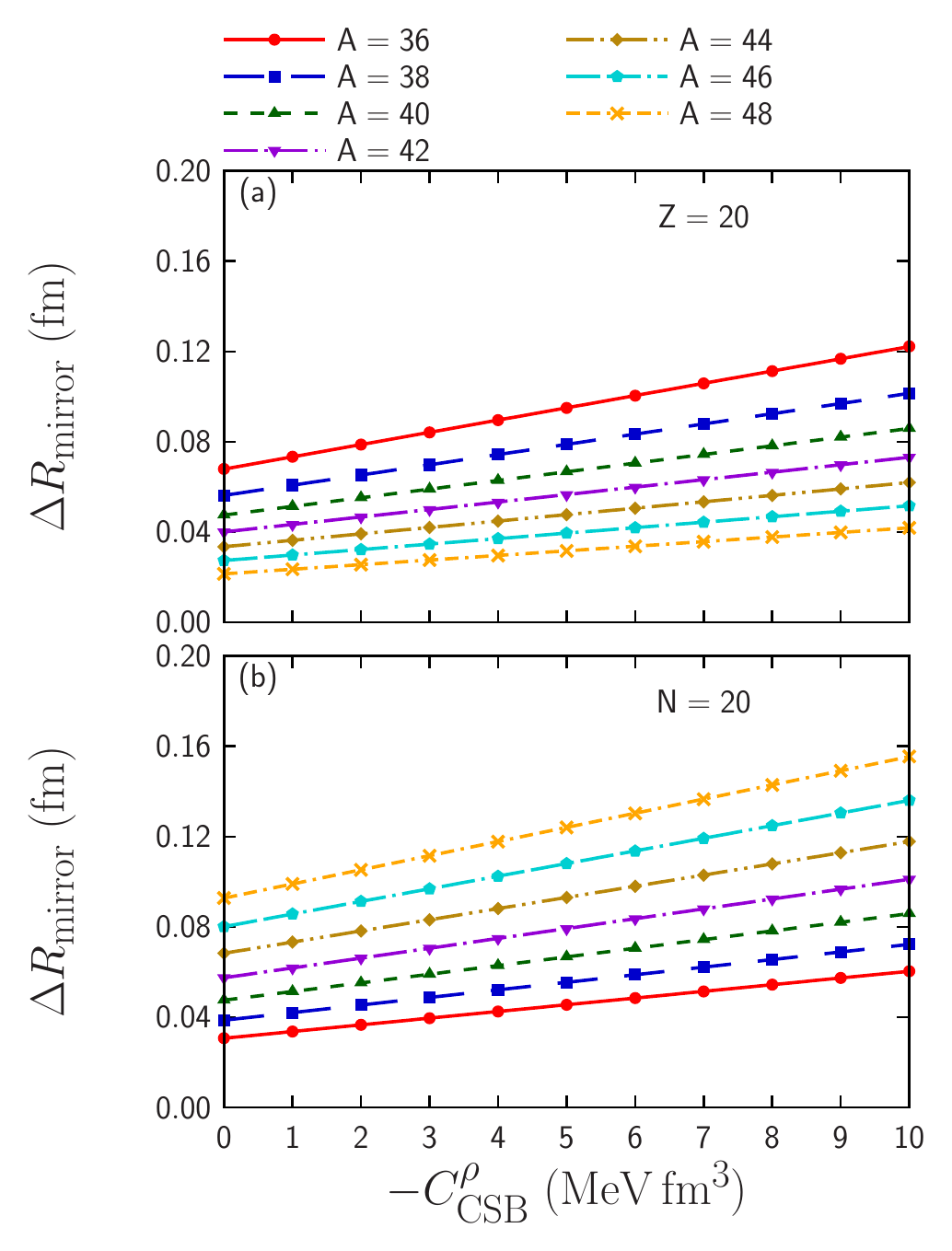}
  \caption{Mirror-skin thickness $ \Delta R_{\urm{mirror}} $ as functions of the CSB strength $ C_{\urm{CSB}}^{\rho} $.
    The SLy4 EDF~\cite{
      Chabanat1998Nucl.Phys.A635_231}
    and 
    the volume-type pairing interaction~\cite{
      Dobaczewski1984Nucl.Phys.A422_103}
    are used for the isospin symmetric particle-hole and particle-particle channel, respectively.
    The Hartree-Fock-Slater approximation, i.e., the local density approximation,
    is used for the Coulomb interaction~\cite{
      Slater1951Phys.Rev.81_385}.
    On top of it, the $ C_{\urm{CSB}}^{\rho} $ term is considered.}
  \label{fig:Crho_Rmirror}
\end{figure}

Figure~\ref{fig:Crho_Rmirror} shows the $ C_{\urm{CSB}}^{\rho} $ dependence of the mirror-skin thickness $ \Delta R_{\urm{mirror}} $.
The $ C_{\urm{CSB}}^{\rho} $ dependence of the mirror-skin thickness is stronger for more proton-rich nuclei,
as expected from Eq.~\eqref{eq:LDM_Rmirror} in Sect.~\ref{sec:mirror_define}.
For instance,
the slopes of $ \nuc{Ca}{36}{} $ ($ 5.4 \times 10^{-3} \, \mathrm{MeV}^{-1} \, \mathrm{fm}^{-2} $)
and
$ \nuc{Ni}{48}{} $ ($ 6.3 \times 10^{-3} \, \mathrm{MeV}^{-1} \, \mathrm{fm}^{-2} $)
are almost three times larger than
that of $ \nuc{Ca}{48}{} $ ($ 2.0 \times 10^{-3} \, \mathrm{MeV}^{-1} \, \mathrm{fm}^{-2} $).
\par
Experimentally, neutron radii can be measured within a $ 0.5 \, \% $ accuracy;
hence, the absolute value of the expected experimental error of $ R_n $ is about $ 0.02 \, \mathrm{fm} $~\cite{
  Zenihiro2010Phys.Rev.C82_044611,
  Zenihiro2018_}.
Proton radii can be extracted from charge radii, which can be measured more accurately.
Therefore, the absolute value of the experimental error of $ \Delta R_{\urm{mirror}} $ can be  about $ 0.02 \, \mathrm{fm} $
and accordingly, the absolute error of $ C_{\urm{CSB}}^{\rho} $ can be $ 6 \, \mathrm{MeV} \, \mathrm{fm}^3 $
using $ \nuc{Ca}{42}{} $.
\subsubsection{$ C_{\urm{CSB}}^{\tau} $- and $ C_{\urm{CSB}}^{\laplace \rho} $-term dependence}
\par
Next, we discuss the $ C_{\urm{CSB}}^{\tau} $ and $ C_{\urm{CSB}}^{\laplace \rho} $ dependences of
$ \Delta R_{\urm{mirror}} $.
To determine the range of $ C_{\urm{CSB}}^{\tau} $,
we compare the contributions to the equation of state by the $ C_{\urm{CSB}}^{\rho} $ term
and by the $ C_{\urm{CSB}}^{\tau} $ term.
Note that the $ C_{\urm{CSB}}^{\laplace \rho} $ term does not contribute to the equation of state.
The CSB contribution to the equation of state reads
\begin{equation}
  \label{eq:EoS}
  \frac{E_{\urm{CSB}}}{A}
  =
  C_{\urm{CSB}}^{\rho}
  \rho
  +
  \frac{4}{5}
  \left( \frac{3 \pi^2}{2} \right)^{2/3}
  C_{\urm{CSB}}^{\tau}
  \rho^{5/3}.
\end{equation}
Therefore, the $ C_{\urm{CSB}}^{\rho} $-term contribution to the equation of state at the saturation density $ \rho_0 $ is
equivalent to the $ C_{\urm{CSB}}^{\tau} $-term under the assumption
\begin{equation}
  C_{\urm{CSB}}^{\tau}
  =
  \frac{5}{4}
  \left( \frac{3 \pi^2}{2} \right)^{-2/3}
  C_{\urm{CSB}}^{\rho}
  \rho_0^{-2/3}.
\end{equation}
For instance, $ C_{\urm{CSB}}^{\rho} \simeq -10 \, \mathrm{MeV} \, \mathrm{fm}^3 $
corresponds to $ C_{\urm{CSB}}^{\tau} \simeq -7.0 \, \mathrm{MeV} \, \mathrm{fm}^5 $
as far as the EoS contribution is concerned.
Considering this relation, we vary $ C_{\urm{CSB}}^{\tau} $ from $ 0 $ to $ -10 \, \mathrm{MeV} \, \mathrm{fm}^5 $.
In contrast, $ C_{\urm{CSB}}^{\laplace \rho} $ can be both negative and positive
because $ \laplace \rho \left( \ve{r} \right) $ can be both negative and positive,
in contrast to $ C_{\urm{CSB}}^{\rho} $ and $ C_{\urm{CSB}}^{\tau} $ dependences;
therefore, we vary $ C_{\urm{CSB}}^{\laplace} $ from $ -5 $ to $ 5 \, \mathrm{MeV} \, \mathrm{fm}^5 $.
\par
\begin{figure}[tb]
  \centering
  \includegraphics[width=1.0\linewidth]{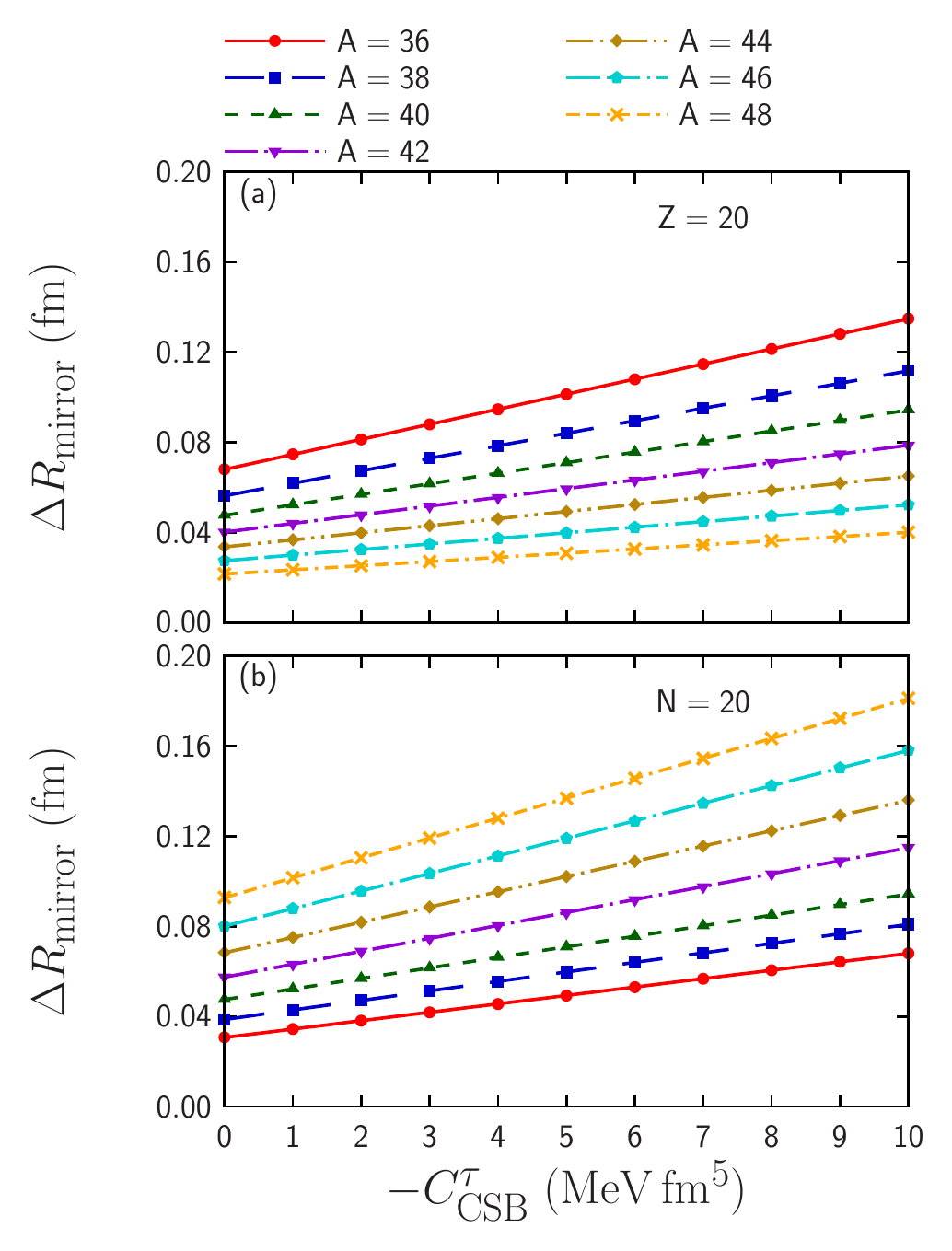}
  \caption{Same as Fig.~\ref{fig:Crho_Rmirror} but for the $ C_{\urm{CSB}}^{\tau} $ dependence.}
  \label{fig:Ctau_Rmirror}
\end{figure}
\begin{figure}[tb]
  \centering
  \includegraphics[width=1.0\linewidth]{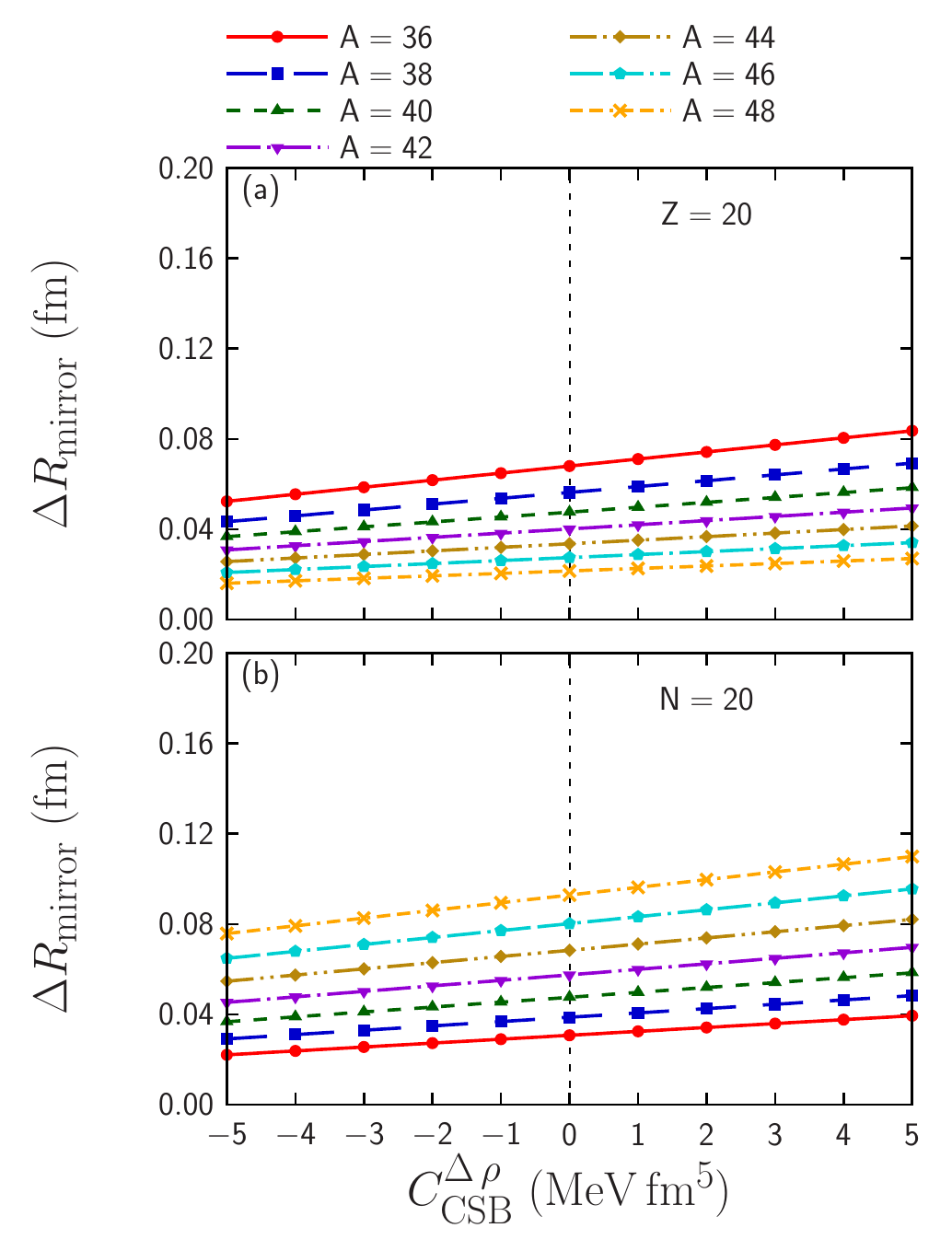}
  \caption{Same as Fig.~\ref{fig:Crho_Rmirror} but for the $ C_{\urm{CSB}}^{\laplace \rho} $ dependence.}
  \label{fig:Clap_Rmirror}
\end{figure}

Figures~\ref{fig:Ctau_Rmirror} and \ref{fig:Clap_Rmirror}, respectively, show the $ C_{\urm{CSB}}^{\tau} $ and $ C_{\urm{CSB}}^{\laplace \rho} $ dependences of $ \Delta R_{\urm{mirror}} $.
It is clearly seen that the $ C_{\urm{CSB}}^{\tau} $ dependence is much stronger than the $ C_{\urm{CSB}}^{\laplace \rho} $ dependence in most cases.
In contrast, the $ C_{\urm{CSB}}^{\tau} $ dependence is similar to the $ C_{\urm{CSB}}^{\rho} $ one.
\par
To understand the reason why the $ C_{\urm{CSB}}^{\tau} $ dependence is the strongest,
we choose the pair of mirror nuclei $ \nuc{Ca}{42}{} $ and $ \nuc{Ti}{42}{} $ as an example.
Hereinafter, for simplicity,
the specific CSB interaction with
$ C_{\urm{CSB}}^{\rho} = -5 \, \mathrm{MeV} \, \mathrm{fm}^3 $ only
($ C_{\urm{CSB}}^{\tau} = -5 \, \mathrm{MeV} \, \mathrm{fm}^5 $ only 
or
$ C_{\urm{CSB}}^{\laplace \rho} = +5 \, \mathrm{MeV} \, \mathrm{fm}^5 $ only)
is 
referred to as 
$ C_{\urm{CSB}}^{\rho} $-CSB
($ C_{\urm{CSB}}^{\tau} $-CSB or $ C_{\urm{CSB}}^{\laplace \rho} $-CSB).
The results are summarized in Table~\ref{tab:ISB_Cparam_Ca42-Ti42}.
\par
\begin{figure}[tb]
  \centering
  \includegraphics[width=1.0\linewidth]{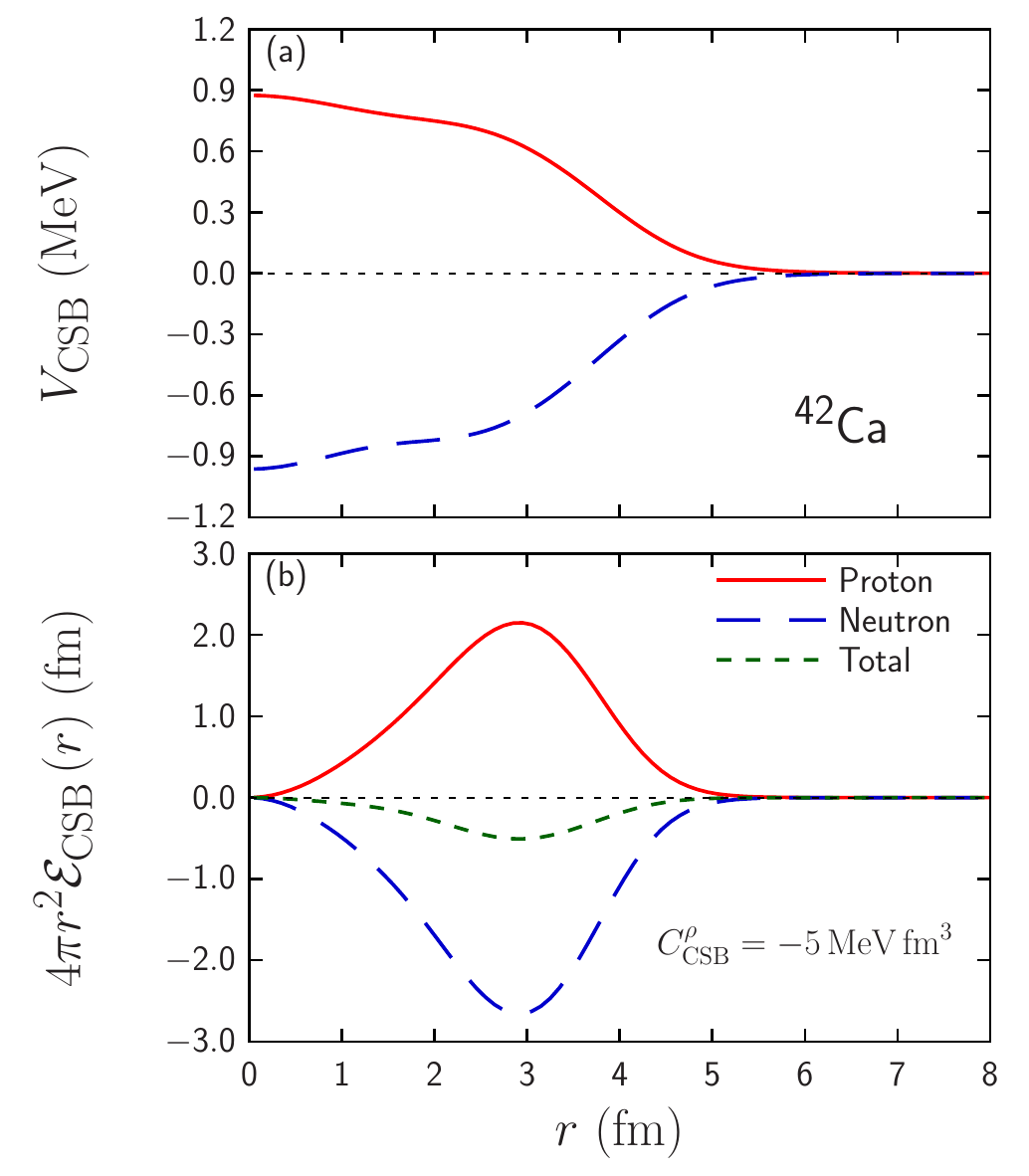}
  \caption{(a) CSB potential and (b) the integrand of the CSB energy density
    of $ \nuc{Ca}{42}{} $,  as functions of $ r $, 
    for $ C_{\urm{CSB}}^{\rho} = -5 \, \mathrm{MeV} \, \mathrm{fm}^3 $.}
  \label{fig:pot_den_Crho_020_042}
\end{figure}
\begin{figure}[tb]
  \centering
  \includegraphics[width=1.0\linewidth]{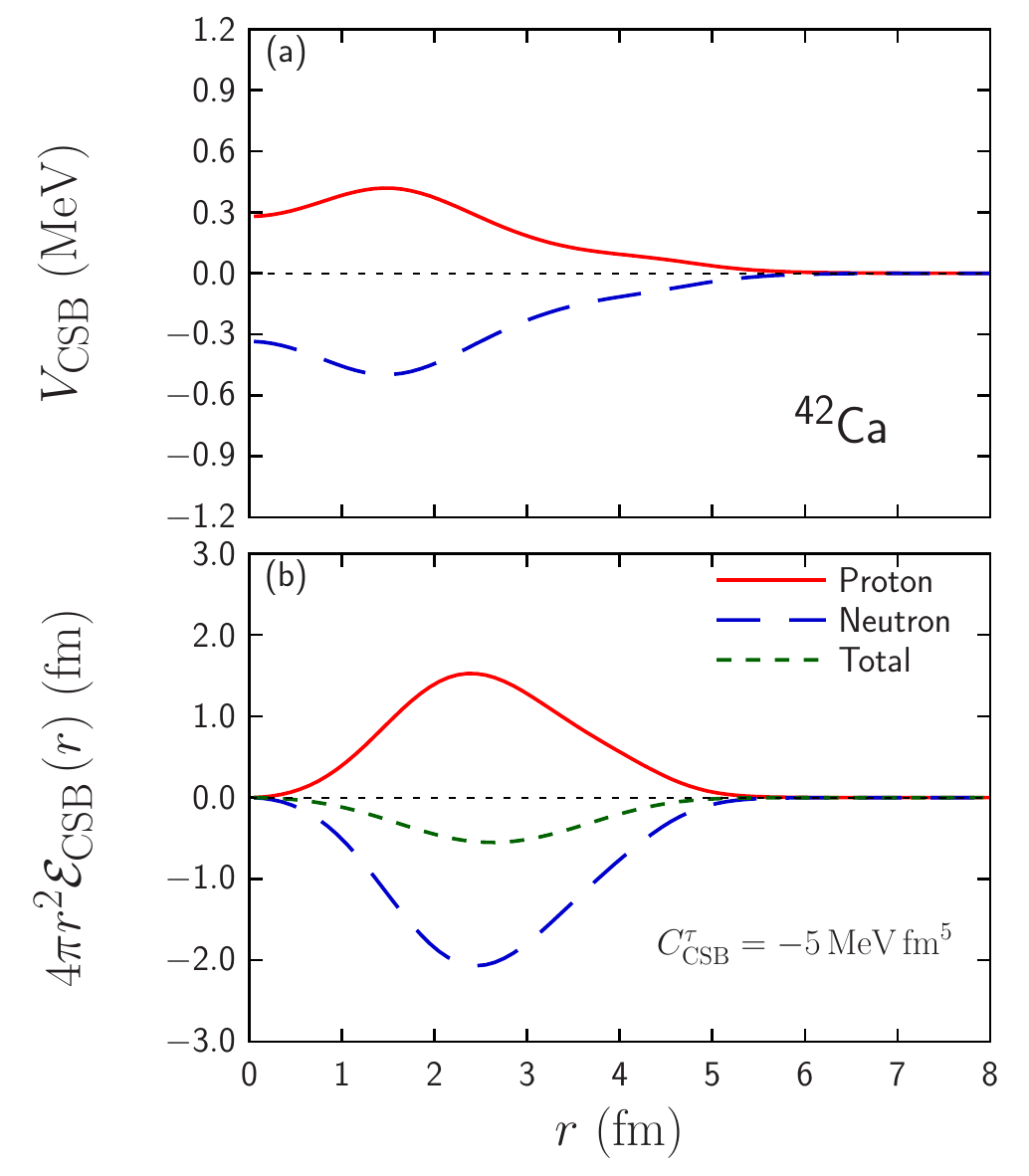}
  \caption{Same as Fig.~\ref{fig:pot_den_Crho_020_042}
    but for $ C_{\urm{CSB}}^{\tau} = -5 \, \mathrm{MeV} \, \mathrm{fm}^5 $.}
  \label{fig:pot_den_Ctau_020_042}
\end{figure}
\begin{figure}[tb]
  \centering
  \includegraphics[width=1.0\linewidth]{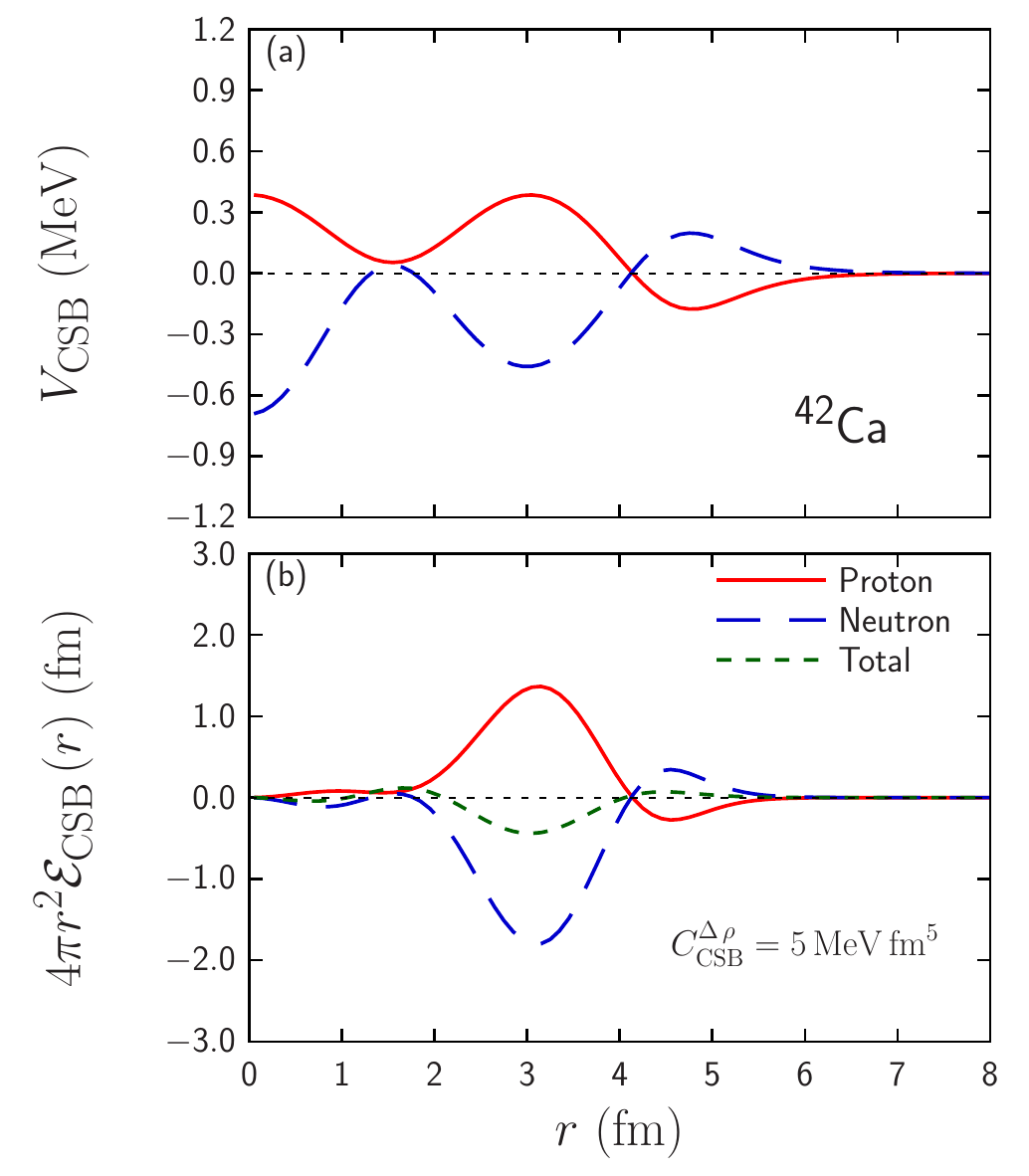}
  \caption{Same as Fig.~\ref{fig:pot_den_Crho_020_042}
    but for $ C_{\urm{CSB}}^{\laplace \rho} = 5 \, \mathrm{MeV} \, \mathrm{fm}^5 $.}
  \label{fig:pot_den_Clap_020_042}
\end{figure}
\begin{figure}[tb]
  \centering
  \includegraphics[width=1.0\linewidth]{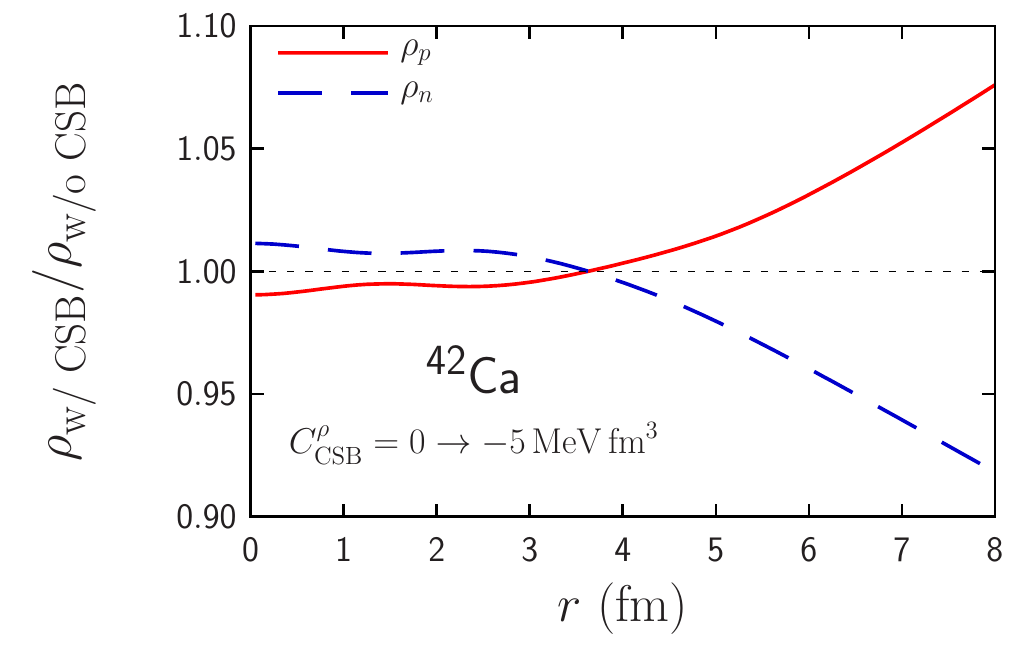}
  \caption{Ratio between the proton and neutron densities of $ \nuc{Ca}{42}{} $,
    $ \rho_p $ and $ \rho_n $, 
    without the CSB interaction
    and those with the CSB interaction ($ C_{\urm{CSB}}^{\rho} = -5 \, \mathrm{MeV} \, \mathrm{fm}^3 $).}
  \label{fig:ratio_den_Crho_020_042}
\end{figure}

First, we shall focus on $ \nuc{Ca}{42}{} $.
Figures~\ref{fig:pot_den_Crho_020_042}--\ref{fig:pot_den_Clap_020_042} show the CSB potential $ V_{\urm{CSB}} $ and the CSB energy density $ 4 \pi r^2 \ca{E}_{\urm{CSB}} $
for $ \nuc{Ca}{42}{} $ with the $ C_{\urm{CSB}}^{\rho} $-CSB, $ C_{\urm{CSB}}^{\tau} $-CSB, and $ C_{\urm{CSB}}^{\laplace \rho} $-CSB interactions.
The CSB energy density in the $ C $-parametrization is given in Eq.~\eqref{eq:CSB_EDF-C},
and the CSB effective mean-field potentials for protons and neutrons, respectively, read
\begin{subequations}
  \begin{align}
    V_{\urm{CSB}}^p \left( \ve{r} \right)
    & =
      \frac{\partial \ca{E}_{\urm{CSB}}}{\delta \rho_p}
      \notag \\
    & =
      -
      \left[
      2 C_{\urm{CSB}}^{\rho}
      \rho_p \left( \ve{r} \right)
      +
      C_{\urm{CSB}}^{\tau}
      \tau_p \left( \ve{r} \right)
      +
      2 C_{\urm{CSB}}^{\laplace \rho}
      \laplace \rho_p \left( \ve{r} \right)
      \right], \\
    V_{\urm{CSB}}^n \left( \ve{r} \right)
    & =
      \frac{\partial \ca{E}_{\urm{CSB}}}{\delta \rho_n}
      \notag \\
    & =
      +
      \left[
      2 C_{\urm{CSB}}^{\rho}
      \rho_n \left( \ve{r} \right)
      +
      C_{\urm{CSB}}^{\tau}
      \tau_n \left( \ve{r} \right) 
      + 
      2 C_{\urm{CSB}}^{\laplace \rho}
      \laplace \rho_n \left( \ve{r} \right)
      \right].
  \end{align}
  \label{eq:CSB_potential}
\end{subequations}
As seen in Eq.~\eqref{eq:CSB_potential}, 
the CSB potential induced by the $ C_{\urm{CSB}}^{\rho} $-CSB interaction is proportional to the proton or neutron density, $ \rho_p $ or $ \rho_n $,
and affects the whole $ \rho_p $ and $ \rho_n $ (see Fig.~\ref{fig:ratio_den_Crho_020_042}).
Eventually, $ \Delta R_{\urm{mirror}} $ increases by about $ 40 \, \% $.
Since $ V_{\urm{CSB}} $ for the neutrons is attractive, $ \rho_n $ shrinks
and since $ V_{\urm{CSB}} $ for the protons is repulsive, $ \rho_p $ extend.
In total, $ \ca{E}_{\urm{CSB}} \left( r \right) r^2 $ is always negative, and thus, $ E_{\urm{CSB}} $ is also negative.
\par
\begin{figure}[tb]
  \centering
  \includegraphics[width=1.0\linewidth]{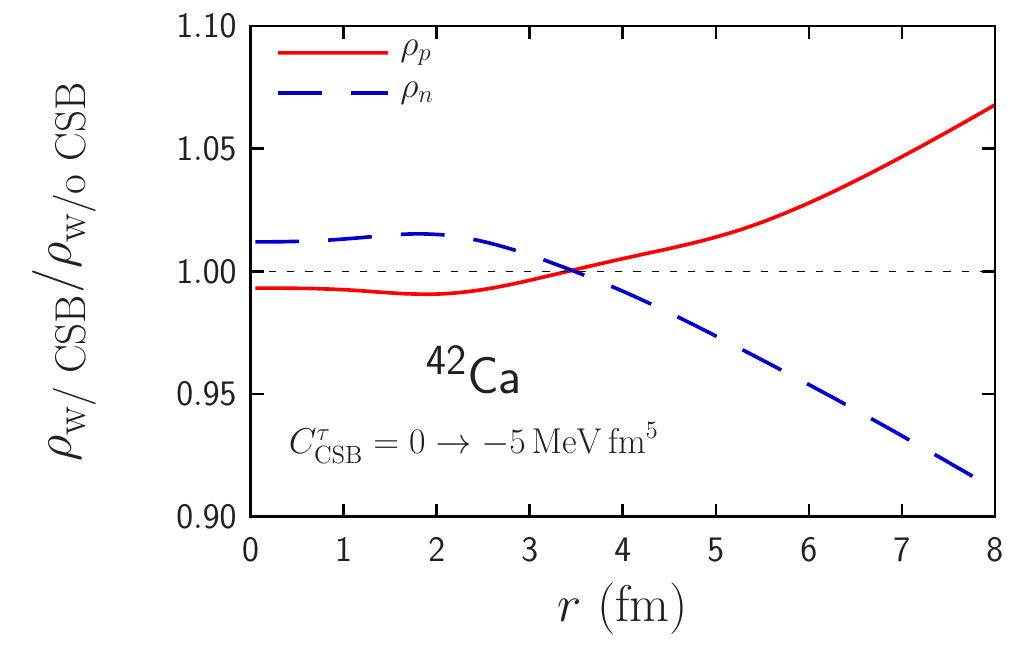}
  \caption{Same as Fig.~\ref{fig:ratio_den_Crho_020_042}
    but for $ C_{\urm{CSB}}^{\tau} = -5 \, \mathrm{MeV} \, \mathrm{fm}^5 $.}
  \label{fig:ratio_den_Ctau_020_042}
\end{figure}

The $ C_{\urm{CSB}}^{\tau} $-CSB interaction shows a similar behavior as the $ C_{\urm{CSB}}^{\rho} $-CSB one.
The $ C_{\urm{CSB}}^{\tau} $-CSB potential is proportional to the kinetic density, $ \tau_p $ or $ \tau_n $,
and accordingly, 
it also affects the whole radial dependence $ \rho_p $ and $ \rho_n $ (see Fig.~\ref{fig:ratio_den_Ctau_020_042}).
Compared to the $ C_{\urm{CSB}}^{\rho} $-CSB potential,
the $ C_{\urm{CSB}}^{\tau} $-CSB potential is weak;
however, the total effect is slightly larger due to the effective mass.
Indeed, the $ C_{\urm{CSB}}^{\tau} $-CSB interaction contributes to the effective mass as
\begin{subequations}
  \begin{align}
    \frac{\hbar^2}{2m^{*p}_{\urm{CSB}} \left( \ve{r} \right)}
    & =
      -
      C_{\urm{CSB}}^{\tau}
      \rho_p \left( \ve{r} \right), \\
    \frac{\hbar^2}{2m^{*n}_{\urm{CSB}} \left( \ve{r} \right)}
    & =
      +
      C_{\urm{CSB}}^{\tau}
      \rho_n \left( \ve{r} \right).
  \end{align}
\end{subequations}
In the case of $ C_{\urm{CSB}}^{\tau} < 0 $,
the proton effective mass becomes lighter, while the neutron one becomes heavier.
Thus, proton radii increase and neutron radii shrink;
accordingly, the mirror-skin thickness becomes larger due to the change in the effective mass.
The change of $ \hbar^2 / 2m^* $ is about
$ 0.5 \, \mathrm{MeV} \, \mathrm{fm}^2$ for $ C_{\urm{CSB}}^{\tau} = -5 \, \mathrm{MeV} \, \mathrm{fm}^5 $,
which is $ 2.5 \, \% $ of the bare mass contribution.
The energy density $ \ca{E}_{\urm{CSB}} \left( r \right) r^2 $ also behaves similarly to the $ C_{\urm{CSB}}^{\rho} $-CSB interaction,
while the absolute value is slightly larger and accordingly the absolute value of $ E_{\urm{CSB}} $ and the CSB effect on $ \Delta R_{\urm{mirror}} $ are also larger by about $ 20 \, \% $.
\par
\begin{figure}[tb]
  \centering
  \includegraphics[width=1.0\linewidth]{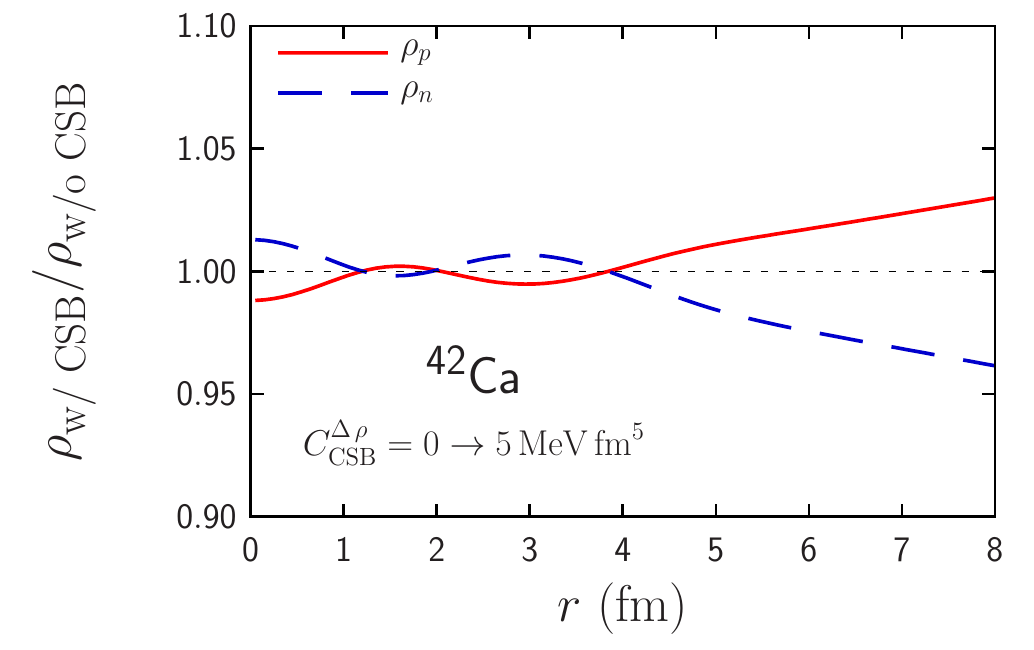}
  \caption{Same as Fig.~\ref{fig:ratio_den_Crho_020_042}
    but for $ C_{\urm{CSB}}^{\laplace \rho} = 5 \, \mathrm{MeV} \, \mathrm{fm}^5 $.}
  \label{fig:ratio_den_Clap_020_042}
\end{figure}

In contrast, the $ C_{\urm{CSB}}^{\laplace \rho} $-CSB interaction behaves differently from the other two CSB interactions.
The $ C_{\urm{CSB}}^{\laplace \rho} $-CSB potential is proportional to $ \laplace \rho_p $ or $ \laplace \rho_n $;
accordingly it has an oscillating structure.
Thus, the net change of $ \rho_p $ and $ \rho_n $ is smaller than in the case of the other two CSB interactions (see Fig.~\ref{fig:ratio_den_Clap_020_042}).
The energy density $ \ca{E}_{\urm{CSB}} \left( r \right) r^2 $ also oscillates and,
accordingly, the absolute value of $ E_{\urm{CSB}} $ and the CSB effect of $ \Delta R_{\urm{mirror}} $ are the smallest.
\par
\begin{figure}[tb]
  \centering
  \includegraphics[width=1.0\linewidth]{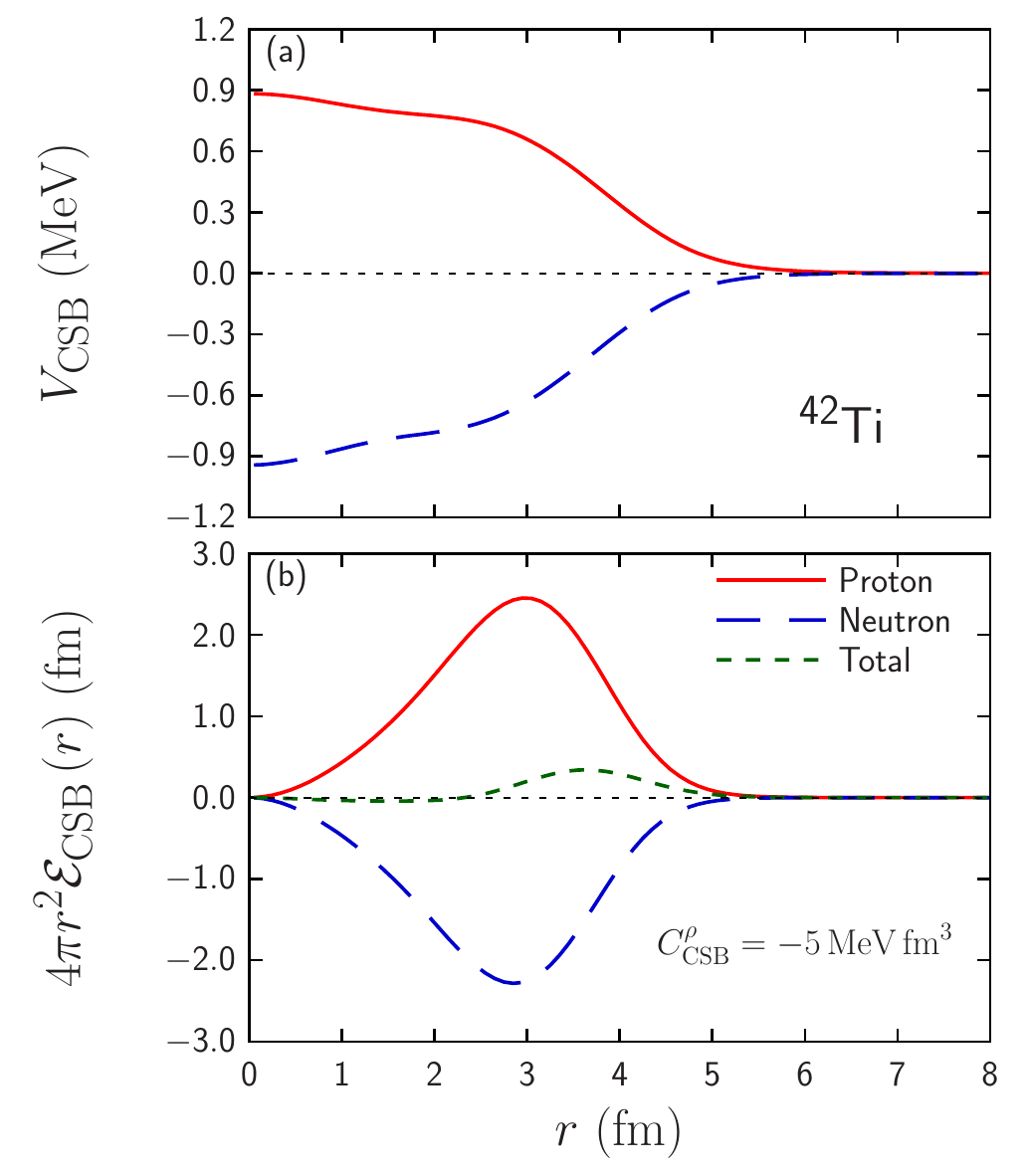}
  \caption{Same as Fig.~\ref{fig:ratio_den_Crho_020_042}
    but for $ \nuc{Ti}{42}{} $.}
  \label{fig:pot_den_Crho_022_042}
\end{figure}

Next, we compare the result of $ \nuc{Ti}{42}{} $ to that of $ \nuc{Ca}{42}{} $.
Here, we take the $ C_{\urm{CSB}}^{\rho} $-CSB interaction as an example.
Figure~\ref{fig:pot_den_Crho_022_042} shows the $ C_{\urm{CSB}}^{\rho} $-CSB potential and the energy density.
The CSB mean-field potential for $ \nuc{Ti}{42}{} $ has a quite similar behavior to that for $ \nuc{Ca}{42}{} $;
and accordingly, the changes of $ R_p $ and $ R_n $ for $ \nuc{Ti}{42}{} $ are almost the same size as those for $ \nuc{Ca}{42}{} $.
However, the energy density almost vanishes in the internal region,
and accordingly, the absolute value of $ E_{\urm{CSB}} $ of $ \nuc{Ti}{42}{} $ is smaller than that of $ \nuc{Ca}{42}{} $.
This difference is basically due to the Coulomb interaction.
Without the Coulomb interaction, $ \rho_p \left( r \right) $ is smaller than $ \rho_n \left( r \right) $ for $ \nuc{Ca}{42}{} $.
Due to the Coulomb interaction, $ \rho_p $ expands, and accordingly,
$ \rho_n^2 - \rho_p^2 $ is always positive.
In contrast, for $ \nuc{Ti}{42}{} $,
without the Coulomb interaction, $ \rho_p \left( r \right) $ is basically larger than $ \rho_n \left( r \right) $.
Due to the Coulomb interaction, $ \rho_p $ extends, and accordingly,
$ \rho_n^2 - \rho_p^2 $ in the internal region reaches to zero.
\par
If we assume the accuracy of $ \Delta R_{\urm{mirror}} $ is about $ 0.02 \, \mathrm{fm} $,
the absolute error of $ C_{\urm{CSB}}^{\tau} $ and $ C_{\urm{CSB}}^{\laplace \rho} $ can be, respectively,
$ 5.2 $ and $ 10.8 \, \mathrm{MeV} \, \mathrm{fm}^3 $
using $ \nuc{Ca}{42}{} $.
\begingroup
\squeezetable
\begin{table*}[tb]
  \centering
  \caption{Total energy $ E_{\urm{tot}} $,
    the CSB energy $ E_{\urm{CSB}} $,
    proton and neutron root-mean-square radii $ R_p $ and $ R_n $,
    and the mirror-skin thickness $ \Delta R_{\urm{mirror}} $
    for $ \nuc{Ca}{42}{} $ and $ \nuc{Ti}{42}{} $ 
    calculated with several ISB parameters and the SLy4 interaction.
    The total energy and the CSB one are shown in $ \mathrm{MeV} $ and
    the others are in $ \mathrm{fm} $.}
  \label{tab:ISB_Cparam_Ca42-Ti42}
  \begin{ruledtabular}
    \begin{tabular}{ddddddddddddd}
      \multicolumn{3}{c}{ISB} & \multicolumn{5}{c}{$ \nuc{Ca}{42}{} $} & \multicolumn{5}{c}{$ \nuc{Ti}{42}{} $} \\
      \cline{1-3}
      \cline{4-8}
      \cline{9-13}
      \multicolumn{1}{c}{$ C_{\urm{CSB}}^{\rho} $} & \multicolumn{1}{c}{$ C_{\urm{CSB}}^{\tau} $} & \multicolumn{1}{c}{$ C_{\urm{CSB}}^{\laplace \rho} $} & \multicolumn{1}{c}{$ E_{\urm{tot}} $} & \multicolumn{1}{c}{$ E_{\urm{CSB}} $} & \multicolumn{1}{c}{$ R_p $} & \multicolumn{1}{c}{$ R_n $} & \multicolumn{1}{c}{$ \Delta R_{\urm{mirror}} $} & \multicolumn{1}{c}{$ E_{\urm{tot}} $} & \multicolumn{1}{c}{$ E_{\urm{CSB}} $} & \multicolumn{1}{c}{$ R_p $} & \multicolumn{1}{c}{$ R_n $} & \multicolumn{1}{c}{$ \Delta R_{\urm{mirror}} $} \\
      \hline
      0  &  0 &  0 & -364.5843 & +0.0000 & 3.4240 & 3.4369 & 0.0401 & -350.2891 & +0.0000 & 3.4944 & 3.3838 & 0.0575 \\
      -5 &  0 &  0 & -365.6038 & -1.0596 & 3.4320 & 3.4260 & 0.0566 & -349.7843 & +0.4587 & 3.5053 & 3.3754 & 0.0793 \\
      0  & -5 &  0 & -365.8122 & -1.3014 & 3.4328 & 3.4219 & 0.0593 & -349.7819 & +0.4215 & 3.5080 & 3.3734 & 0.0861 \\
      0  &  0 & +5 & -364.9563 & -0.3949 & 3.4284 & 3.4306 & 0.0494 & -350.1905 & +0.0748 & 3.5003 & 3.3789 & 0.0697 \\
    \end{tabular}
  \end{ruledtabular}
\end{table*}
\endgroup
%
%
\section{Summary}
\label{sec:summary}
\par
In this paper, we proposed a new observable, named the mirror-skin thickness $ \Delta R_{\urm{mirror}} $,
to pin down the magnitude of the CSB term in the nuclear EDF.
As examples, we studied the mirror-skin thickness of $ N = 20 $ isotones and $ Z = 20 $ isotopes by using HFB calculations with various Skyrme EDFs and adding CSB and CIB terms.
It is shown that the mirror-skin thickness is sensitive only to the CSB EDF, but hardly depends on either the isospin symmetric part of the nuclear interaction or the CIB term.
Therefore, this observable can be used to determine the CSB EDF quantitatively either from experimental data or \textit{ab initio} calculation. 
\par
The $ \rho^2 $ and $ \rho \tau $ terms of the CSB EDF give similar substantial contributions to $ \Delta R_{\urm{mirror}} $,
while the $ \rho \laplace \rho $ one gives the smallest contribution.
Therefore, it is possible to extract the values of $ C_{\urm{CSB}}^{\rho} $ or $ C_{\urm{CSB}}^{\tau} $,
namely the coupling constants of the $ \rho^2 $ and the $ \rho \tau $ terms,  
using experimental information on the mirror-skin thickness,
while it is difficult to obtain the $ C_{\urm{CSB}}^{\laplace \rho} $ value. 
Among the mirror pairs,
we propose the experimental study of the mirror-skin thickness between $ \nuc{Ca}{42}{} $ and $ \nuc{Ti}{42}{} $,
or even more exotic pairs of mirror nuclei,
which could be accessed in future experiments at RIBF and/or FRIB.
The current accuracy on measurement of neutron radii could be enough to extract the information on the CSB interaction,
while more accurate measurement and the measurement of density profiles can provide more accurate determination of the coupling constants.
\par
As far as the present study and several previous studies in the literature~\cite{
  Baczyk2018Phys.Lett.B778_178,
  Baczyk2019J.Phys.G46_03LT01,
  Naito2022Phys.Rev.C106_L061306,
  Naito2023Phys.Rev.C107_064302}
are concerned, the CSB interaction always gives stronger contribution to most observables than the CIB interaction.
Thus, it is an important future perspective to find observables that are equally sensitive, or more sensitive, to the CIB interaction than to the CSB interaction.
One possible future direction may be using triplet of isobars as it was done for masses in Refs.~\cite{
  Baczyk2018Phys.Lett.B778_178,
  Baczyk2019J.Phys.G46_03LT01}.
References~\cite{
  Sato2013Phys.Rev.C88_061301,
  Sheikh2014Phys.Rev.C89_054317}
proposed more complete treatments of the isospin considering the proton-neutron mixed densities,
which may help us towards the complete study on the isospin symmetry breaking.
%
%
\begin{acknowledgments}
  T.~N.~acknolwedges
  the RIKEN Special Postdoctoral Researcher Program,
  the JSPS Grant-in-Aid for Research Activity Start-up under Grant No.~JP22K20372,
  the JSPS Grant-in-Aid for Transformative Research Areas (A) under Grant Nos.~JP23H04526 and JP25H01558,
  the JSPS Grant-in-Aid for Scientific Research (S) under Grant No.~JP25H00402,
  the JSPS Grant-in-Aid for Scientific Research (B) under Grant Nos.~JP23H01845, JP23K26538, and JP25K01003,
  the JSPS Grant-in-Aid for Scientific Research (C) under Grant No.~JP23K03426,
  the JSPS Grant-in-Aid for Early-Career Scientists under Grant No.~JP24K17057,
  and
  the JSPS Grant-in-Aid for JSPS Fellows under Grant No.~JP25KJ0405.
  The numerical calculations were performed on cluster computers at the RIKEN iTHEMS program.
\end{acknowledgments}
\clearpage
\appendix
\section{Dependence on modelling of the Coulomb interaction of mirror-skin thickness}
\label{sec:Coul}
\par
In this appendix, we show the mirror-skin thickness does not depend on the modelling of the Coulomb interaction.
As done in Ref.~\cite{
  Naito2020Phys.Rev.C101_064311},
we test several modelling of the Coulomb interaction.
The Coulomb Hartree approximation (without the exchange term),
the Coulomb Hartree-Fock local density approximation (Coulomb Hartree-Fock-Slater approximation),
and
the Coulomb Hartree-Fock generalized gradient approximation (GGA)
are, respectively, referred to as ``NoCx'', ``LDA'', and ``GGA''.
On top of the Coulomb GGA calculation,
electric form factors of protons, neutrons, and the vacuum polarization are gradually considered,
which are referred to as ``$ p $-fin'', ``$ pn $-fin'', and ``All'', respectively.
\par
In this appendix, the pairing is not considered and thus Skyrme Hartree-Fock calculation is performed
using the \textsc{skyrme\_rpa} code~\cite{
  Colo2013Comput.Phys.Commun.184_142}
with the SLy4 interaction and $ C_{\urm{CSB}}^{\rho} = - 3.2875 \, \mathrm{MeV} \, \mathrm{fm}^3 $,
which is the suggested value in Ref.~\cite{
  Roca-Maza2018Phys.Rev.Lett.120_202501}.
The mirror-skin thickness for $ \nuc{Ca}{48}{} $ and $ \nuc{Ni}{48}{} $ are shown in Table~\ref{tab:Coul}.
It is found that the modelling dependence of the mirror-skin thickness is about $ O \left( 0.001 \right) \, \mathrm{fm} $
as long as the Coulomb exchange term is considered,
while the CSB effect is about $ 0.05 \, \mathrm{fm} $.
Therefore, we can safely neglect the uncertainty originating from the modelling of the Coulomb interaction
as long as the Coulomb exchange term is considered.
\begin{table}[!htb]
  \centering
  \caption{Dependencing on modelling of the Coulomb interaction of mirror-skin thickness of $ \nuc{Ca}{48}{} $ and $ \nuc{Ni}{48}{} $
    calculated with the SLy4 interaction.}
  \label{tab:Coul}
  \begin{ruledtabular}
    \begin{tabular}{ldddd}
      Model & \multicolumn{2}{c}{Without CSB ($ \mathrm{fm} $)} & \multicolumn{2}{c}{With CSB ($ \mathrm{fm} $)} \\
            & \multicolumn{1}{c}{$ \nuc{Ca}{48}{} $} & \multicolumn{1}{c}{$ \nuc{Ni}{48}{} $} & \multicolumn{1}{c}{$ \nuc{Ca}{48}{} $} & \multicolumn{1}{c}{$ \nuc{Ni}{48}{} $} \\
      \hline
      NoCx       & 0.0246 & 0.1019 & 0.0379 & 0.1434 \\
      LDA        & 0.0215 & 0.0927 & 0.0350 & 0.1338 \\
      GGA        & 0.0215 & 0.0922 & 0.0348 & 0.1333 \\
      $ p $-fin  & 0.0197 & 0.0862 & 0.0332 & 0.1271 \\
      $ pn $-fin & 0.0198 & 0.0863 & 0.0332 & 0.1272 \\
      All        & 0.0199 & 0.0870 & 0.0333 & 0.1279 \\
    \end{tabular}
  \end{ruledtabular}
\end{table}
%
%
%
%
%
\end{document}